\documentstyle[epsf,12pt]{article}

\def\be		{\begin{equation}}
\def\ee		{\end{equation}}
\def\Tr		{\mathop{\hbox{Tr}}}
\def\la		{\left\langle}
\def\ra		{\right\rangle}
\def\max	{\hbox{\tiny max}}
\def\tot	{\hbox{\tiny tot}}
\def\sec	{\hbox{\tiny 2nd}}
\def\diff	{\hbox{\tiny diff}}
\def\eff	{\hbox{\tiny eff}}
\def\uone	{\hbox{\tiny U(1)}}
\def\mon	{\hbox{\tiny mon}}
\def\sw		{\hbox{\tiny sw}}
\def\arts	{\sqrt K}
\def\artsq	{K}

\newlength{\figsize}
\begin{document}

\begin{titlepage}

\begin{tabbing}
\` {\sl hep-lat/9712003} \\
    \\
\` LSUHE No. 268-1997 \\
\` OUTP-97-64P \\
\` November, 1997 \\
\` revised March, 1998 \\
\end{tabbing}
 
\vspace*{1.0in}
 
\begin{center}
{\large\bf Monopole clusters in Abelian projected gauge theories\\}
\vspace*{.5in}
A. Hart$^1$ and M. Teper$^2$\\
\vspace*{.2in}
{\it $^1$Department of Physics and Astronomy, Louisiana State University,\\ 
Baton Rouge, LA 70803, U.S.A.\\
e--mail: hart@rouge.phys.lsu.edu}
\\
\vspace*{.2in}
{\it $^2$Theoretical Physics, University of Oxford,\\
1 Keble Road, Oxford, OX1 3NP, U.K.\\
e--mail: teper@thphys.ox.ac.uk}
\end{center}

\vfill


\end{titlepage}
\vfill\eject

\pagestyle{empty}

\begin{center}
{\bf Abstract}
\end{center}

We show that the monopole currents which one obtains in the maximally
Abelian gauge of SU(2) fall into two quite distinct classes (when the
volume is large enough). In each field configuration there is
precisely one cluster that permeates the whole lattice volume. It has
a current density and a magnetic 
screening mass that scale and it
produces the whole of the string tension.  The remaining clusters have
a number density that follows an approximate power law $\propto
{1\over{l^3}}$ where $l$ is the length of the monopole world line in
lattice units. These clusters are localised in space-time with radii
which vary as $\surd l$. In terms of the radius $r$ these `lumps' have
a scale-invariant distribution $\propto {{dr}\over r} \times {1 \over
{r^4}}$.  Moreover they appear not to contribute at all to the string
tension. The fact that they are scale-invariant at small distances
would seem to rule out an instanton origin.

\setcounter{page}{0}
\newpage
\pagestyle{plain}

\section{Introduction}

Magnetic monopole currents are the crucial degrees of freedom in the
dual superconducting vacuum hypothesis for confinement in non-Abelian
gauge theories
\cite{mandelstam76,thooft81}.
After Abelian projection to the maximally Abelian gauge
\cite{thooft81,kronfeld87a},
one finds not only that the Abelian fields possess a string tension,
$\sigma$, that (almost) equals the original SU(2) string tension
(`Abelian dominance')
\cite{suzuki90},
but that this string tension is almost entirely due to the monopoles
in those Abelian fields (`monopole dominance')
\cite{stack94,bali96}.
If the dual superconductor hypothesis is indeed correct, then the
magnetic monopoles reflect that part of the infrared physics in the
SU(2) vacuum which drives confinement. It is therefore of great
interest to analyse the structure of the monopole currents so as to
determine whether there are any simple or suggestive features present.
This is our goal in this paper.

We shall focus on some simple properties of these monopoles.  Our
basic tool is to decompose the total monopole current into
non-intersecting clusters. An alternative would be to decompose the
current into closed loops; for example a monopole cluster might be
decomposed into several closed loops that intersect. There is no
obvious reason why the monopole cluster spectrum should be more
revealing than the loop spectrum, and indeed in an earlier study
\cite{hart97a}
we have found that this loop spectrum does possess some interesting
features. As we shall see below, however, it turns out that it is the
cluster spectrum that possesses the simplest and most remarkable
properties.

In the next section we briefly discuss the technical details of the
calculation, including the Abelian projection, the extraction of the
string tension and the parameters of the lattice
simulations. Section~\ref{mon_conf} contains a simple analytic
calculation showing how monopoles can cause Abelian Wilson loops to
decay exponentially with their area.  The purpose of this simplistic
but useful picture is to give some orientation as to what properties
the monopoles must possess if they are to be confining. This enables
us to motivate bounds on the type of monopole spectrum that can be
confining.  

In section~\ref{sec_infrared} we present the evidence for our most
striking result: that the monopole current contains a single
`percolating' cluster that permeates the whole volume, together with a
collection of smaller clusters, whose number density, as a function of
length, $l$, is close to an inverse cubic power. Such a spectrum
decays slowly enough with increasing $l$ that it can in principle
confine. Our explicit calculations show, however, that it makes no
contribution to the string tension, within errors, and that it is the
single largest cluster that provides the string tension. We then
analyse the scaling properties of these clusters. We show that the
length per unit volume of the largest cluster is constant when
expressed in physical units. This is not so for the remaining
clusters. 
We find that at large distances, $r$, from a monopole the magnetic
flux falls exponentially with $r$ and that the
corresponding screening mass is independent of the lattice spacing. 
If, however, we calculate the flux at smaller values of $r$,
where the flux is large enough to efficiently disorder appropriately
positioned Wilson loops, we find that scaling is violated except 
if we only include the monopoles that belong to the largest cluster.  
As a
further tool we introduce a method for locally smoothing the monopole
currents. This shows us that the fact that we have a {\em single} huge
cluster must have a dynamical origin rather than being a simple
`percolation' phenomenon. Also we see that the largest cluster
possesses substantial fluctuations that do not add to its confining
properties. 

The smaller clusters are typically localised within a 4-volume whose
radius $r\propto \surd l$. We find that these `4-balls' possess a
scale invariant distribution, $\propto {{dr}\over r} \times {1 \over
{r^4}}$. If the scale-invariance of the gauge theory were not
anomalous, then this is precisely the distribution one would have for
instantons. Given that we know that instantons are associated with
monopole loops within their cores
\cite{hart96,chernodub95}
this would have provided an elegant explanation. Unfortunately the
anomalous breaking of scale-invariance changes the instanton spectrum
in a dramatic and calculable fashion for the small values of $r$ where
the spectrum of the `4-balls' is most accurately determined. Thus this
seems to rule out instantons as being relevant. 

In Section~\ref{sec_summary} we provide a summary of our results and some
conclusions. A brief summary of some of our results has appeared in
\cite{hart97c}.
We draw the reader's attention to some related work that has appeared
recently
\cite{fukushima97}.
\section{Methodology}
\label{sec_method}

The first step in our calculation is to generate SU(2) lattice field
configurations. We use the standard Wilson plaquette action and a
standard heat bath Monte Carlo algorithm.  The lattices have periodic
boundary conditions. We work with $8^4$, $10^4$, $12^4$ and $16^4$
lattices at $\beta=2.3$, with $10^4$, $12^4$ , $14^4$ and $16^4$
lattices at $\beta=2.4$, and with $16^4$ lattices at $\beta=2.5$. The
range of lattice sizes at fixed $\beta$ is intended to provide us with
control over finite volume effects. For example, $16a\sim 6 \times
1/\surd\sigma$ at $\beta=2.3$: a very large length in units of the
physical length scale. The range of $\beta$ values is intended to
provide us with some control over finite-$a$ corrections ($a$
decreases by about a factor of 2 between $\beta=2.3$ and
$\beta=2.5$). We typically analyse 500 configurations for each $L$ and
$\beta$. These configurations are typically some 25 to 50 Monte Carlo
sweeps apart.

Once generated these SU(2) configurations are then fixed to the
maximally Abelian gauge in the standard way: we perform gauge
transformations at each site, and iterate the procedure, so as to
(locally) maximise the gauge dependent functional
\be
R = - \sum_{n,\mu} \Tr \left( 
U_\mu(n).i\sigma_3.U^\dagger_\mu(n).i\sigma_3 
\right).
\label{A1}
\ee
We then write the gauge fixed links in the factored form
\be
U_\mu(n) = 
\left(
\begin{array}{cc}
c_\mu(n) 	& w_\mu(n) \\
-w^\ast_\mu(n)	& c_\mu(n)
\end{array}
\right)
\left(
\begin{array}{cc}
e^{i \theta_\mu(n)} & 0 \\
0 & e^{-i \theta_\mu(n)}
\end{array}
\right),
\label{A2}
\ee
where $c_\mu(n)$ is real and the $\theta_\mu(n)$ are our Abelian link
angles. We now identify the magnetic monopole currents in these
Abelian fields using
\cite{degrand80}.
The currents are integer valued variables on the links of the dual
lattice and they satisfy a continuity equation. So the total current
can be decomposed into a number of closed current loops. In general
such a decomposition is not unique since loops may intersect. If loops
that intersect are concatenated into `clusters' then these clusters
form a unique set of mutually disconnected networks and each current
link may be unambiguously associated with one of these clusters.

A standard way to calculate the SU(2) string tension is by calculating
Wilson loops, $W(r,t)$: i.e. the trace of the oriented product of
SU(2) matrices along the rectangular $r \times t$ contour.  From these
Wilson loops one can extract the static potential, $V(r)$:
\be
aV(r) = \lim_{t \to \infty} 
\ln \left\{ \frac{\la W(r,t) \ra}{\la W(r,t+a) \ra} \right\}.
\label{A3}
\ee
>From the behaviour of $V(r)$ at large $r$, $V(r) \sim \sigma r$, we
can then extract the string tension, $\sigma$. Clearly such a
calculation, involving two limits, requires large lattices and small
errors. An alternative procedure is to use Creutz ratios:
\be
a^2 \sigma = \lim_{r \to \infty} \sigma_{\eff}(r) 
\equiv - \lim_{r \to \infty} 
\ln \left\{ {{\la W(r,r) \ra \la W(r+a,r+a) \ra }\over{\la W(r+a,r) \ra 
\la W(r,r+a) \ra }} \right\}.
\label{A4}
\ee
In practice a useful estimate of the string tension can be extracted 
this way when the quality of the `data' does not permit the
preceding, more complete analysis.

Once we have gauge fixed and extracted our Abelian fields, we can
obtain the Abelian string tension in exactly the same way.  We simply
calculate the Wilson loops using the Abelian fields $u_\mu (n) =
\exp\{ i\theta_\mu (n)\}$ rather than the SU(2) matrices $U_\mu
(n)$. The fact that this Abelian string tension turns out to be close
to the full SU(2) string tension
\cite{suzuki90},
has provided much of the motivation for the current interest in
the maximally Abelian gauge.

To calculate the monopole contribution to a Wilson loop let us
consider contours that are purely space-like e.g. $W(x,y)$.  (Since
space-time is Euclidean, this involves no loss of generality.)  The
integral of the Abelian gauge potential around the contour will simply
equal the magnetic flux, $B(x,y)$, through a surface spanning the
Wilson loop contour, so the value of the Abelian Wilson loop will be
given by
\be
W(x,y) = \exp [i B(x,y)].
\label{A5}
\ee
In principle the surface chosen should be one over which the Abelian
potential is non-singular. But since the flux through any other surface
will differ by an integer multiple of $2\pi$ (Dirac strings),
we are free to choose whichever surface is the most convenient ---
which will usually be the minimal surface.  The monopole Wilson loop
is obtained by using that part of the magnetic flux that is generated
by the monopole charges. This is just the dual of the electric flux
that would be generated by the corresponding electric charges. We
calculate this flux by solving the dual Maxwell equations with the
given monopole currents. This is done by an iterative procedure and
for the particular periodic four-volume under consideration. Once one
has the dual 4-potential, it is trivial to generalise the calculation
to non-space-like Wilson loops.  Calculating Wilson loops in this way
we can extract the monopole potential and string tension, using
eqns.~(\ref{A3}) or~(\ref{A4}).

In the same way one can, if one wishes, calculate the string tension
due to some specified subset of monopole clusters. One simply
calculates the dual potential due to that subset of monopole currents.

If we were working with a U(1) theory then we would expect the whole
of the Abelian string tension to be due to monopoles
\cite{polyakov77,banks77}.
In the present case, however, the Abelian fields are not generated by
a (semi-)local Abelian action but are obtained in a complicated way
from the non-Abelian fields. It is therefore possible that the
resulting vacuum contains confinement-inducing, disordering
fluctuations other than monopoles. For example, if the vacuum were to
contain finite-width tubes of magnetic flux, with the flux, say, equal
to $\pi$, and if these loops were to be arbitrarily long (a
`condensate') then this would typically produce a non-zero string
tension.  Thus it is important to ask whether it is the case that
within the ensemble of Abelian fields obtained by Abelian projection
from the SU(2) fields, confinement is indeed generated entirely by
monopoles.  A first step is to calculate both the Abelian and monopole
string tensions and to compare them. Several investigations of this
type suggest that they are indeed quite similar
\cite{stack94,bali96}.
To go further than this we need to directly compare the confining
Abelian and monopole fluctuations. To do this we calculate on each
field configuration the difference between the total magnetic flux and
that due to the monopoles.  Using this `difference' flux we then
calculate the corresponding Wilson loops and potentials. If the string
tension that we extract from this potential is zero, then we will have
shown that the confining fluctuations in the Abelian fields are
entirely due to the monopoles. We have performed such calculations and
display a typical set of results in Table~\ref{tab_sigma_diff}. The
effective string tension, $\sigma_{\eff}(r)$, has been obtained from
Creutz ratios, as in eqn.~(\ref{A4}). We observe that within errors
the `difference' string tension is indeed consistent with being
zero. This provides direct evidence that confinement is entirely
driven by monopoles in these U(1) fields.

The reader will note something rather peculiar about the numbers in
Table~\ref{tab_sigma_diff}. It is apparent that the monopole Creutz
ratios show very much smaller statistical fluctuations than those from
the U(1) fields.  We would therefore expect that the difference string
tension should have statistical errors that are at least as large as
those in the U(1) measurement. In fact, as we see, they are much
smaller.  This clearly requires a strong correlation between the
fluctuations in the U(1) fields and in the monopole currents: as we
expect to be the case from monopole dominance. This is not in itself
sufficient to explain the pattern of fluctuations, however.  We note
also that the small-$r$, Coulombic deviation of the difference
potential away from the purely linear asymptotic form is much greater
than in the pure monopole calculation. This is actually something we
can rather easily understand, as we shall see in
section~\ref{sec_infrared}.

Finally two cautionary asides. The first concerns Gribov copies.  The
gauge fixing described above is not unique. The gauge functional has
many maxima: the well known Gribov copy problem. These copies are, of
course, identical for gauge invariant quantities but differ for gauge
variant quantities such as the Abelian fields and monopole
currents. Since there is currently no convincing criterion for which
maximum is the `best', we shall simply ignore this ambiguity. A
practical justification for doing so is the demonstration
\cite{hart97b}
that while the monopole content of different Gribov copies of the same
SU(2) gauge field can be very different, the long distance monopole
physics that produces confinement is in fact very similar for typical
Gribov copies. Since confinement is what we are mainly interested in
here, this reassures us that our qualitative conclusions should not be
affected by the Gribov copy problem.

The second aside concerns positivity. Since the SU(2) action is local
(i.e. it extends over only a fixed number of lattice units) there is a
corresponding positive-definite Hamiltonian in the continuum limit and
it makes sense to talk of masses, potentials etc. (For non-zero
lattice spacing there might be peculiar effects for masses on the
order of the cut-off.) There is no guarantee, however, that the
ensemble of Abelian fields possesses such an underlying Hamiltonian
--- because the Abelian fields depend in a completely non-local manner
on the original SU(2) fields --- and so we cannot be certain that it
makes sense to talk of Abelian potentials and string tensions. This
applies even more so to the ensemble of monopole fields; and even more
to situations where we consider only subsets of monopole currents.
Having said this, one finds in practice that the Abelian and monopole
Wilson loops usually do behave as if there were an underlying transfer
matrix, and the extraction of the potential seems to be largely
unambiguous. So we will follow previous work and ignore possible
problems with positivity. That these problems do exist becomes
immediately apparent if one tries to `modernise' the calculation using
smearing/blocking techniques. The correlation functions of smeared
operators badly break positivity.  This undermines the usual
variational approach and means that we can only be confident that we
have obtained the lightest mass if we have a clear, extended effective
mass plateau.  These problems have occasionally arisen in our
calculations, but not in those that are reported upon in this paper.

\section{Monopoles and confinement}
\label{mon_conf}

Before moving on to our results concerning the distribution of
monopole currents it is interesting to ask whether there are any
restrictions or bounds that such a distribution should satisfy if it
is to have any possibility of producing confinement.  To do so it will
first be useful to outline how monopoles produce confinement in
Abelian theories. The focus here will be on identifying the essential
features of the phenomenon and will involve a variety of simplifying
approximations to the exact calculations
\cite{polyakov77,banks77}.
\subsection{A simple picture}

To begin with we consider the simpler case of the 3-dimensional U(1)
theory. Here the monopoles are really instantons, but because the
fields are identical to time-sliced fields from the static
4-dimensional U(1) theory, it is customary and appropriate to refer to
them as monopoles and to talk of the fields as being
magnetic. Suppose, then, that we consider a Wilson loop on an $R
\times T$ contour. The contribution of the monopoles to the value of
the Wilson loop is just
\be
\la W(R,T) \ra  = \la \exp [i B_{\mon}(R,T)] \ra 
\label{A6}
\ee
where the average is over all field configurations and $B_{\mon}(R,T)$
is the total magnetic flux through the $R \times T$ rectangle that
arises from the monopoles in each field configuration.

How do we calculate $B_{\mon}(R,T)$? One might try to neglect the
monopole interactions as a first approximation, so that we just have a
random gas of monopoles. This leads to arbitrarily large energy
densities, however, and so the system prefers to trade off some
entropy and form a screened plasma of magnetic charges instead. Let
the screening length be $\xi$. We shall treat our system as being, to
a first approximation, a random gas of monopoles with a screened
magnetic flux that decreases with distance $r$ as $b_{\pm}(r) = \pm
2\pi \exp(-r/\xi)$ (the sign being chosen at random). Consider now the
total flux $\Phi$ through our $R \times T$ contour. Given the exponential
drop in the flux a reasonable approximation for $R,T \gg \xi$ is to
assume that if a monopole lies within a `slab' of thickness $\xi$
either side of the $R\times T$ rectangle then half of its flux,
i.e. $\Phi = \pi$, will pass through the rectangle while if it is
outside the slab then the flux is suppressed to zero. This obviously
neglects various perimeter effects, but we do not care because these
will not contribute to the interesting area term. In this
approximation then
\be
B_{\mon}(R,T) = n_{+} \pi -  n_{-} \pi
\label{A7}
\ee
where $n_{+}$ ($n_{-}$) is the number of positively (negatively)
charged monopoles above the Wilson rectangle plus the number of
negatively (positively) charged monopoles below --- counting only
those within the slab of course.  Clearly once $R,T \gg \xi$ the
average number of monopoles within the slab must be proportional to
its volume $\la n_{+} \ra = \la n_{-} \ra = c \xi R T \equiv {\bar
n}$.  If the gas in the slab is random then $\la n_{\pm} \ra $ should
be Poisson distributed with mean $\bar n$. We can now calculate our
Wilson loop average:
\begin{eqnarray}
\la W(R,T) \ra  &=& \la \exp [i B_{\mon}(R,T)] \ra  \nonumber \\
&=& \sum_{n_{+}=0} e^{i n_{+} \Phi} {{{\bar n}^{n_{+}}}\over{n_{+}!}} 
e^{-{\bar n}}
\times
\sum_{n_{-}=0} e^{-i n_{-}\Phi} {{{\bar n}^{n_{-}}}\over{n_{-}!}} 
e^{-{\bar n}} \nonumber \\
&=& e^{-2 {\bar n}\left( 1- \cos \Phi \right) } \nonumber \\
&=& e^{-4c\xi RT}
\label{A8}
\end{eqnarray}
using $\Phi=\pi$ in the last line.
Thus the monopole magnetic flux causes the Wilson loops to decay
exponentially with the loop area. This means that the monopoles lead
to a non-zero string tension: $\sigma = 4c\xi$, in the above
approximation.

The mechanism here is very simple. Only a monopole within a distance
$\xi$ will significantly affect the Wilson loop because of
screening. Its contribution to the flux is about $\Phi \sim \pi$ and
so it flips the sign of the loop: $e^{i\pi} = -1$. That is to say,
these monopoles maximally disorder the loop. Their number is obviously
proportional to the area and this immediately translates into an area
decay and a corresponding string tension.

We can easily do a bit better. If we consider a monopole a distance $r$
above a large Wilson loop, the screened flux through that loop is
\be
\Phi(r) = \pi \int_0^1 dy . e^{-\frac{r}{y\xi}}.
\ee
This assumes that $\xi \ll R,T$ so that we are only interested
in $r \ll R,T$, in which case the flux through the Wilson loop
is (almost) the same as the flux through the whole plane in
which the loop lies. Using our previous expression for the
average screened flux, we readily obtain the above $\Phi(r)$.  
Multiplicatively combining the disordering effects of an infinite
tower of slabs, each infinitesimally thick, we obtain by analogy
to eqn.~(\ref{A8}):
\be
\la W(R,T) \ra  = e^{-2c\xi RT \int_{0}^{\infty} dr
\left \{ 1 - \cos \left(\pi q \int_0^1 dy e^{- {r \over y}} 
\right) \right\}}
\label{A9}
\ee
We have now introduced a general electric charge $q$ for the Wilson
loop. Since we have chosen the magnetic charge to be unity the usual
Dirac quantisation relation tells us that $q$ must be an integer
(otherwise the Dirac strings become `visible'). For $q=\pm 1$
eqn.~(\ref{A9}) is an inessential improvement.  For, say, $q=2$,
however, the argument of the previous paragraph gives no confinement
since a flux of $\pi$ does not disorder a doubly charged Wilson loop:
$e^{\pm i2\pi} = 1$. So in this case it is eqn.~(\ref{A9}) that must
used and we obtain a string tension
\be
\sigma (q) = 2 c \xi \int_{0}^{\infty} dr
\left\{ 1 - \cos \left( \pi q \int_0^1 dy e^{- {r \over y}} \right) \right\}
\label{A10}
\ee
for the potential between static sources of charge $q$.

It is crucial, if we are to obtain confinement, that screening is
something that occurs only on the average --- it is a statistical
phenomenon. If, for example, we were to consider a gas of magnetic
dipoles --- a non-statistical form of screening --- then we would get
no confinement: the net flux through our very large Wilson loop is
essentially zero if the dipole is well within the perimeter of the
loop and a distance $\ll R,T$ from the surface of the loop.  The fact
that screening is statistical means that the fluctuations around the
mean screened flux will be important. The Wilson loop is of course
sensitive to all fluctuations --- it is, after all, a phase --- and so
we are making an uncontrolled approximation in replacing the monopole
fluxes by their mean, screened values. This is the only serious
approximation that we have made. We shall return to the link between
the confining properties of the monopoles, the monopole current
density and the screening length in the next section.

Although our treatment of screening is very approximate, this does not
undermine the simple picture we gave above of how monopoles maximally
disorder Wilson loops, and so maximise the interaction between
electric charges.  Indeed suppose we ignore screening entirely and
calculate an $R\times R$ Wilson loop, say, within a completely random
gas of monopoles. The calculation is now actually much easier since
there is no screening length to bring in an extra scale.  We can
therefore just scale out the scale $R$ and we obtain
\be
\la W(R,R) \ra  \propto e^{-c R^3}
\label{A11}
\ee
where $c$ is proportional to the density of monopoles.  The cubic
power of $R$ arises on simple dimensional grounds. So the potential
grows faster than linearly: a random gas of monopoles over-confines.
This is not possible in a quantum field theory: the associated
unbounded energy densities will break down through particle
production. Screening is the way the theory regulates itself and in
the process weakens the over-confining potential to the linear form
that is the fastest growth possible for a proper field theory
\cite{seiler78}.

The above simple and, no doubt, well-known picture contains the
essential features of how monopoles drive linear confinement in 3
dimensions and, for static monopoles, in 4 dimensions as well. Of
course we are interested here in the non-static case. Since space-time
is Euclidean we lose no generality by considering only space-like
Wilson loops. In that case it continues to be the magnetic flux that
disorders the Wilson loop, exactly as above. It is still the case that
the net magnetic flux from a monopole will be $2\pi$. Of course this
flux will no longer be spherically symmetric but will depend on the
motion of the monopoles. The generic effect of this asymmetry is to
weaken the string tension but only by a finite factor that should not
be far from unity on the average. Thus the qualitative physics is
unchanged. If we time-slice monopole loops that are much smaller than
our Wilson loop, they will look like dipoles and will not disorder the
Wilson loop sufficiently to confine. The same should apply to Wilson
loops that are long in one direction but short in another. The
qualitative conclusion is that confinement on a scale $R$, requires
monopole loops that are large compared to the corresponding $R\times
R$ Wilson loops. (A numerical confirmation of this may be found in
\cite{stack94}.
where it is seen that small monopole loops do not contribute to the
string tension.) One could try to go further but we shall stop here
and see what we can infer from this rather general constraint.

\subsection{Bounds on a confining monopole spectrum}

So we now ask what conditions $N(l)$, the number of clusters of length
$l$, must satisfy if we are to have confinement. We shall take the
lattice spacing to be fixed so that the only quantity we vary is the
lattice volume: $L^4$ in lattice units.
 
We start with the simplifying assumption that the monopole cluster
spectrum, $N(l)$, falls off as a power of $l$:
\be
N(l) = \frac{C(L)}{l^\gamma}.
\label{A12}
\ee
Our arguments can be straightforwardly adapted to other functional
forms but we choose to focus on a power law because we already know
that the spectrum of monopole loops decreases approximately as $\sim
1/l^3$
\cite{hart97a}.
Moreover, as we shall see in the next section, the cluster spectrum
also possesses such a component. Once the volume is large compared to
the physical length scale, we expect the $L$-dependence of $C(L)$ to
be simply $C(L)=c L^4$. The first bound then arises if we make the
reasonable assumption that the density of monopole current must be
finite, i.e.
\be
\lim_{L\to\infty}
{{c L^4 \int  l {{dl}\over{l^\gamma}}}
\over {4 L^4}} \not= \infty.
\label{A13}
\ee
This equation immediately implies that
\be
\gamma \ge 2.
\label{A14}
\ee
as long as the maximum length of those clusters which are
associated with the power law, 
$l_{\max}(L)$, grows $\to \infty$ when $L \to\infty$. 
In general this must be the case. Indeed simple random walk
arguments would suggest that $l_{\max}(L) \propto L^2$.

We now have a lower bound on $\gamma$. Confinement should provide us
with some upper bound: after all if $\gamma$ is large enough then
there will be essentially no large monopole clusters to disorder large
Wilson loops. Let us be more specific. Consider Wilson loops of size
$\epsilon L \times \epsilon L$, on an $L^4$ lattice. If the theory is
confining then it is reasonable to expect that it should be confining
on scales $\epsilon L$ where $\epsilon$ can be chosen arbitrarily
small but is then fixed. This will require monopole clusters that
extend over distances of order $\epsilon L$ at least. Let $L$ be so
large that $\epsilon L$ is large compared to the physical length
scale. Then we expect from random walk arguments that the length of
such a cluster should be at least $\propto (\epsilon L)^2$.  This
should certainly apply to the coarse-grained length (the length of the
cluster after the smallest ultraviolet fluctuations in the current
have been removed by smoothing or blocking up to the physical length
scale). Now, let the fraction of configurations with clusters that are
this long, i.e. $ l \ge (\epsilon L)^2$, be $f(l)$.  Clearly if $f(l)
\to 0$ as $L \to \infty$ then we will have lost confinement on the size 
scale $\sim \epsilon L$.  So we require
\be
\lim_{L \to \infty} f(l) \propto  \lim_{L \to \infty} 
c L^4 \int\limits_{\sim (\epsilon L)^2 } {{dl}\over{l^\gamma}} \ne 0
\label{A15}
\ee
which immediately implies
\be
\gamma \le 3.
\label{A16}
\ee
We note that our discussion assumes, as seems reasonable, that the
clusters are essentially independent of each other, i.e. that there
are no strong long-range correlations between different
clusters. Obviously a highly ordered set of small clusters can
simulate the effects of a large cluster, and this would undermine our
above arguments and bounds.

Thus as long as the monopoles possess some very general physical
properties, the exponent characterising the number density is limited
to the narrow range
\be
2 \le \gamma \le 3
\label{A17}
\ee
if the monopoles are to provide the disordering fluctuations that
drive confinement. By making more specific assumptions one can try to
narrow this range, but one then increasingly relies on arguments of
decreasing plausibility. We shall not pursue this here.

Our above arguments have thus led us to the conclusion that a spectrum
of the form $N(l) \propto 1/l^3$ might be confining. This is
intriguing: one finds just such a distribution for monopole loops
\cite{hart97a}
and, as we shall shortly see, for monopole clusters as well.  Moreover
it has been suggested
\cite{diakonov95}
that for large $l$ such a distribution could arise from instantons.

\section{Infrared behaviour}
\label{sec_infrared}

As described above, for each field configuration we extract the
associated monopole current, $\{ j_\mu(n) \}$. The current is integer
valued and conserved. Therefore it can be decomposed into continuous
closed loops of non-zero current. Such a decomposition is ambiguous
when loops cross. If we now form `clusters' of monopole currents by
saying that two loops belong to the same cluster if and only if they
intersect, then the decomposition into clusters is clearly unique. In
this paper we shall focus on clusters rather than loops.  (For an
investigation of the latter see
\cite{hart97a}.)
In addition to being constrained to form closed loops, the currents
must satisfy a further constraint due to the periodic boundary
conditions.  Periodicity demands that in any given time-slice the
total magnetic charge must be zero. Contractible monopole loops
automatically satisfy this requirement.  A loop can also satisfy
current conservation, however, by closing through one of the
boundaries. Periodicity then requires that such winding loops be
matched by other non-contractible loops so that the net charge is zero
however we time-slice the lattice.  We mention this fact since it will
become important when we attempt to calculate the string tension that
arises from a subset of the clusters.

Suppose we have a particular cluster $C$. Then we define the length of
the cluster to be
\be
l_C = \sum_{\{n,\mu\}\in C} \left| j_\mu(n) \right|.
\label{A18}
\ee
In practice, if one is outside the strong-coupling region of the SU(2)
theory then the current is almost always~$\pm 1$ when it is
non-zero. Thus our definition almost coincides with the number of
links in the cluster.

\subsection{Cluster decomposition}

Our first step is to calculate the length of each cluster.  This
reveals that the clusters fall into two quite distinct classes. First
there is a $single$ cluster that is very much longer than any of the
other clusters (at least if the volume is large enough). For example,
of the 500 $16^4$ configurations that we analysed at $\beta=2.3$ there
was not a single case where we observed two large clusters rather than
just one.  Secondly there are the remaining, smaller clusters. These
possess a spectrum which follows an approximate power-law, $N(l) =
\frac{C(L)}{l^\gamma}$, with $\gamma \simeq 3$.

That the very large cluster is not simply the largest cluster of a
smooth distribution of clusters can be established as follows.  First,
simply at the qualitative level, we note that on the $12^4$ lattice at
$\beta=2.3$ (for example) the average length of the largest cluster is
$\sim 3200$, while the average length of the second largest is only
$\sim 49$. Moreover on none of the 500 analysed configurations is the
second largest cluster ever larger than 220 or the largest cluster
ever smaller than 2300. On $16^4$ the distinction is even more marked,
with the average length of the largest cluster $\sim 10169$, while
that of the second largest is only $\sim 67$. We can of course be less
impressionistic than this. First we remark (see
Table~\ref{tab_lengths}) that the total length of all the clusters,
$l_{\tot}$, is proportional to the volume, $L^4$, as is the length of
the largest cluster, $l_{\max}$.  So, as one would expect, the
normalisation of the spectrum is $C(L) = c L^4$. (One would expect
this, because the number of clusters of length $l$ should be
proportional to the volume once the volume is large enough.) On the
average a field configuration will contain $\int N(l)dl$ clusters that
belong to this spectrum.  The largest of these clusters will, roughly
speaking, be sampled from the tail of $N(l)$ that integrates to unity:
\be
c L^4 \int_{l_0} {{dl}\over{l^\gamma}} = 1
\label{A19}
\ee
>From this we can estimate the average length of this largest 
cluster to be
\be
\la l_1 \ra  = {{\int_{l_0} {{ldl}\over{l^\gamma}}}
\over{\int_{l_0} {{dl}\over{l^\gamma}}}}
\propto L^{4 \over{\gamma - 1}}
\label{A20}
\ee
That is to say, because this largest cluster is sampled from a falling
spectrum its length increases much more slowly than $L^4$. This is in
contrast to the observation in Table~\ref{tab_lengths} that the length
of the largest cluster in fact increases as $L^4$. Therefore this
cluster does not belong to the observed continuous spectrum of
clusters. One can also analyse the probability of observing such a
large cluster, if it is sampled from an extrapolation of our observed
spectrum. This probability is negligibly small. Indeed it is the
second largest cluster that appears to be the largest cluster that is
drawn from the continuous part of the spectrum. We list its length as
a function of $L$ in Table~\ref{tab_lengths} and we can see that it
increases weakly with $L$ --- just the behaviour we argued for above.

\subsection{Scaling and the largest cluster}

\begin{figure}
\begin{center}

\leavevmode
\epsfxsize = \figsize
\epsffile{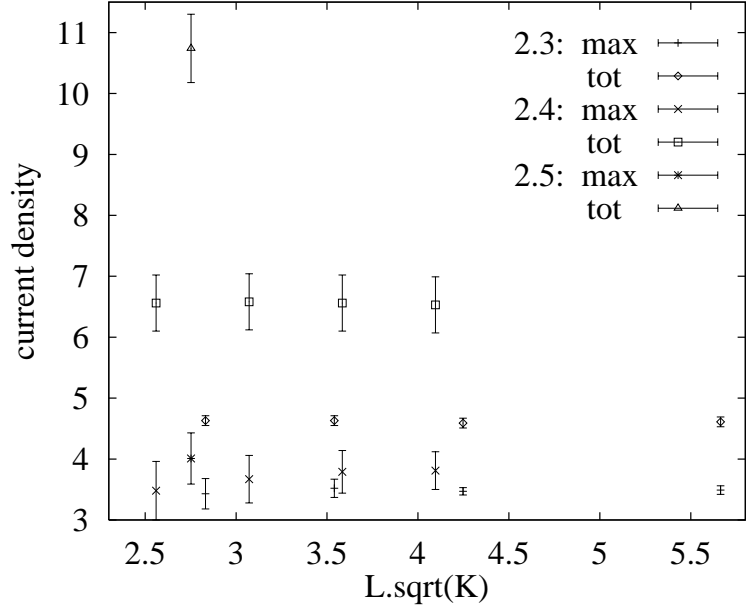}
\end{center}
\caption{
The total current density, $\rho_{\tot}$, and that of the largest 
cluster, $\rho_{\max}$, as functions of the physical lattice size 
and the lattice spacing; for $\beta = 2.3$,~2.4 and ~2.5.}
\label{fig_lmax_scaling}
\end{figure}

We have seen that at fixed $a$ the length of the largest cluster
increases linearly with the volume. This means that it will spread
throughout the space-time volume in the thermodynamic limit. This is,
qualitatively at least, exactly the kind of monopole cluster that
might give us confinement on arbitrarily large scales. If it is to do
so, however, then its structure must encode the physical length scale
and not just the lattice scale. Consider then the length of this
largest cluster in physical units: this will be $l_{\max} \arts$,
where $l_{\max}$ is the length in lattice units and 
$\artsq \equiv a^2\sigma$ is the SU(2) lattice string tension
\cite{michael88}.
Similarly the lattice volume
will be $(L \arts)^4$ in these physical units.
So the monopole current density, in these nonperturbatively 
defined	physical units, is
\be
\rho_{\max} = 
{l_{\max}.\arts \over (L.\arts)^4} = 
{l_{\max} \over L^4 (\arts)^3}.
\label{A21}
\ee
for the largest cluster. Similarly we define $\rho_{\tot}$ for the
total monopole current. We plot these densities against the physical
lattice size, $L \arts$, in Figure~\ref{fig_lmax_scaling}. We first
note that the points at fixed $\beta$ are constant for both
$\rho_{\tot}$ and $\rho_{\max}$. This tells us that for fixed $a$ both
the total current and that from the largest cluster increase linearly
with the volume; something we have noted already.  Comparing now the
points corresponding to different values of $\beta$ we see that
$\rho_{\max}$ is independent of the variation in $a$. That is to say,
the length of the largest cluster is proportional to the volume if
everything is expressed in physical units. By contrast we observe that
this is certainly not the case for the total current. Since a large
part of the total current resides in the largest cluster, this tells
us that there are very strong scaling violations arising from the
currents of the smaller clusters.

The simple scaling property of the largest cluster is quite
remarkable. Of course, it reassures us that the largest cluster does
indeed encode the physical length scale, but it goes further than
that. Realistically one could only hope for a suitably coarse-grained
cluster length to satisfy scaling.  Na\"{\i}vely one would expect the
length defined in terms of the links to be an ultraviolet length whose
relationship to the physical coarse-grained length would involve
anomalous dimensions that would lead to a violation of scaling. This
would arise from the fact that the monopole world line has
fluctuations on all length scales. Instead what we infer is that the
largest monopole cluster does not really seem to encode any
information concerning the ultraviolet length scale.

\subsection{The monopole potential and string tension}

Having seen that the largest cluster fills space-time in a way
that scales in physical units, we now ask whether it does in fact
contribute to confinement, i.e. does it generate a potential between
static sources that has a non-zero string tension.  As described
previously, we can do this by first calculating the magnetic field
arising from the largest monopole cluster and then, from this,
calculating the values of space-like Wilson loops.  (In practice we
calculate the dual 4-potential and all orientations of Wilson loops.)
>From the Wilson loops we extract the monopole potential, $V(r)$, as in
eqn.~(\ref{A3}), and obtain the string tension by fitting it with the
generic form $V(r) = a + b/r + \sigma r$.

First an aside about the $b/r$ term. It has been noted before, e.g.
\cite{hart97a},
that $b$ is very small for the full monopole current ensembles. We
find that the potential from the largest cluster alone is even more
linear in form. Na\"{\i}vely we would expect two contributions
$\propto 1/r$: a Coulomb interaction and the universal L\"uscher
correction to the flux tube energy. Are they both absent or are they
cancelling each other? The latter possibility is not as implausible as
it might at first appear.  We know that in the Villain model Wilson
loops exactly factorise into spin-wave and monopole pieces
\cite{polyakov77,banks77}.
Hence the total potential is a sum: $V(r) = V_{\sw}(r) + V_{\mon}(r)$,
using an obvious notation. Suppose we are in the confining phase. Then
$V(r)$ has a linear piece $\sigma r$ and, in addition, a Coulomb
piece, $V_C = -\alpha/r$, at small $r$ and a L\"uscher term, $V_L= -
\pi/12r$, at large $r$.  These have the same sign and there is no
possibility of a cancellation. In any case, since there is no massless
gluon, the Coulomb piece will be screened at large $r$, typically
$V_C(r) \simeq -{\alpha\over r}\exp\{-{r\over\xi}\}$, and since the
flux tube has a finite width, the L\"uscher term will be `screened' at
small $r$, crudely $V_L(r) \sim
-{\pi\over{12r}}(1-\exp\{-{r\over{\xi^{\prime}}}\})$.  We expect the
two scales, $\xi$ and $\xi^{\prime}$, to be similar, so in practice
the Coulomb and L\"uscher terms will hardly overlap. In contrast to
this the spin-wave potential does possess a massless photon and no
linear piece, i.e. $V_{\sw}(r) \simeq -{\alpha\over r}$ for all
$r$. Therefore $V_{\mon}(r)=V(r)-V_{\sw}(r)$ will have the form
\be
V_{\mon}(r) \simeq c -{\alpha\over r} \exp
\left[ -{r\over\xi} \right]
-{\pi\over{12r}} \left( 1-
\exp \left[ -{r\over{\xi^{\prime}}} \right] \right)
+\sigma r
+{\alpha\over r}.
\label{21b}
\ee
We note that at small $r$ the two Coulomb terms cancel and the
L\"uscher term is negligible: so there is no significant $1/r$
piece. At large $r$ the screened Coulomb term is negligible and so
$V_{\mon}(r)$ will be the difference of the L\"uscher and unscreened
Coulomb terms. If these have a similar magnitude, as in fact they do
in typical SU(2) calculations, then they will largely cancel. So
$V_{\mon}(r)$ has no significant $1/r$ contribution at large $r$
either.  Of course all this has only been demonstrated in the Villain
model. It seems plausible to us, however, that an approximate version
of this mechanism should hold more generally, and that it provides the
explanation for the observed lack of a significant $1/r$ piece in the
monopole potentials.

In calculating the potential from the largest monopole cluster there
is one significant problem. As we remarked earlier periodicity implies
that the total magnetic charge in any time-slice must be zero. So it
needs to be the case that the `pruned' configurations formed by the
removal of certain clusters have no net winding number in any
direction, e.g.
\be
\sum_{x,y,z} j_4(x,y,z,t = 1) = 0
\label{A22}.
\ee
Without this, Gauss' law and the magnetic flux are ill-defined.  In
practice the largest cluster does sometimes have a net winding in one
or more directions. In these cases we implement the following
`fix'. We manually correct the winding number to zero by the addition
of straight lines of magnetic current of appropriate charge that wrap
around the lattice; structures analogous to the Polyakov line of gauge
links. The position of such a static monopole world line is chosen
randomly. In practice this problem is only significant at
$\beta=2.5$. The reason is that the total current necessarily has zero
winding. So the largest cluster will only wind if some other smaller
cluster has a compensating winding. To have some chance of winding
around a lattice this secondary cluster cannot be too small. Only at
$\beta=2.5$ are the secondary clusters large enough to make this a
frequent occurrence.  (This in part arises because of their scaling
properties - as discussed in the next section.) We estimate the bias
induced by our winding fix as follows. We place the same static
monopole lines used to correct the winding number onto otherwise empty
lattices. To each of these, we add a similar number of oppositely
charged lines, also at random positions, and calculate the string
tension. One half of this is an estimate of the bias inherent in our
correction method. (Of course, as we showed earlier, a random monopole
gas over-confines; here we are simply obtaining an effective string
tension at an appropriate distance.)  This bias is found to be
completely negligible at $\beta = 2.3$ and $2.4$. At $\beta = 2.5$ it
is $\pm 0.002$ in lattice units.  This is only an estimate, so the
message is that some caution should be attached to the string tensions
we calculate at $\beta=2.5$.
	
In Table~\ref{tab_sigma_prune} we show the string tensions calculated
using: first the total current, second the largest cluster and third
all the clusters except the largest. We also indicate the proportion
of current in the largest cluster by quoting the value of the
percolation parameter
\cite{hands89}
\be
P = {n_{\max} \over n_{\tot}}.
\label{A23}
\ee
where $n_{\max}$ is the number of (dual) sites connected by current
links from the largest cluster and $n_{\tot}$ the number connected in
all clusters. We note that at $\beta=2.3$, where the volume is largest
in physical units, the string tension is given entirely by the largest
cluster and there is no contribution from the secondary clusters,
despite the fact that the latter carry some $25 \%$ of the total
monopole current.  At $\beta=2.4$ the secondary clusters still do not
provide any confining force even though their contribution to the
current is now approaching half of the total.  At $\beta=2.5$ the
situation is not so clear-cut but that is not surprising: the volume
is now becoming quite small in physical units, 
the distinction between the largest and
second largest clusters begins to blur, and the winding correction has
become important. There is also some indication that
$a^2\sigma_{\tot}$ and $a^2\sigma_{\max}$ --- and indeed
$a^2\sigma_{\tot}+a^2\sigma_{\max}$ --- are not quite the same at
$\beta=2.4$.  We believe that this is directly related to our
observation that as the volume decreases (at fixed $\beta$) there is a
growing disparity between $a^2\sigma_{\tot}$ and $a^2\sigma_{\max} +
a^2\sigma_{\hbox{\tiny rest}}$.  We are not at present sure whether
this indicates a significant correlation between the largest and
smaller clusters on smaller volumes, or whether it is an artifact of
the difficulty of extracting extended effective mass plateaux on small
lattices.

\subsection{Smoothing the monopole fields}

We have just seen that the largest cluster is the source of all the
interesting confining physics. Given its importance it is worth
probing its structure in more detail. In particular we return to our
earlier observation of scaling and the question of whether this
cluster fluctuates on the scale of the lattice spacing or not. To
address this question we locally `smooth' the monopole currents and
see what effect this has on the length of the cluster. We have
employed two methods. The first is simply to `cool' the Abelian fields
by locally changing the fields so as to maximise the value of the sum
of plaquettes.  (That is, we cool using a plaquette `action'.) This
directly smoothes the Abelian fields and therefore, indirectly, the
monopole currents as well. The second method involves `smoothing' the
monopole currents directly: we examine each (dual) plaquette in turn,
and superimpose a $1 \times 1$ current loop with a charge chosen to
minimise the total magnetic current on the lattice. This constitutes
one smoothing sweep. It directly removes the kinks (`staples') in the
current. The two methods give similar results, but the latter one has
the important advantage for us that, in addition to being more
transparent, it enables us to smooth individual clusters if we so
wish.

The result of smoothing the monopole fields, by the second method, is
summarised in Table~\ref{tab_huge_cool}. The first thing we observe is
that the string tension shows very little variation with cooling. This
is as it should be: the ultraviolet fluctuations of the monopole
current should not affect its confining properties. Secondly we note
that, as we cool, there is a rapid decrease in the total length of the
largest cluster. This tells us that it does contain fluctuations on
the size scale of the `cut-off' even if these are not strong enough to
destroy the scaling of the total length. We also find something else
that is very interesting: the largest cluster frequently breaks up
into more than one cluster even after just one smoothing sweep. To see
this we display in Table~\ref{tab_huge_cool} not just the results of
smoothing all the monopole clusters, but also what happens if we
exclude the original largest cluster from consideration and smooth
just the secondary clusters.  We observe that already after just one
smoothing sweep the largest of the latter, labeled $l_{\sec}^\prime$,
is on average much smaller than the second largest cluster, labeled
$l_{\sec}$, obtained when we smooth all the clusters.  (Note that for
these quantities the numbers in brackets are not the errors but the
one standard deviation variations.)  Thus the second largest cluster
must have hived off from the largest cluster. Since $l_{\sec}$
increases with smoothing (initially) it is clear that the largest
cluster is hiving off a substantial number of clusters during the
first few smoothing sweeps.  This raises a puzzle. These hived-off
clusters are, as we can see, typically much larger than the second
largest cluster that one observes prior to smoothing. This implies
that this hiving off never occurs during the Monte Carlo
generation. It certainly would occur if we were applying a Monte Carlo
directly to the U(1) or monopole fields --- after all the smoothing is
just a particular move that would be part of the Monte Carlo choice.
The implication appears to be that there is something in the SU(2)
dynamics that ensures the presence of just one large monopole cluster.
Despite initial appearances this largest cluster cannot be understood
as a simple U(1) monopole percolation phenomenon.

\subsection{Screening lengths}

As we saw in section~\ref{mon_conf}, a plasma of monopoles
is confining and the resulting string tension is proportional 
to the product of the monopole density and the screening length.
Since we have found that the ensemble of monopole currents
that we generate by going to the maximally Abelian gauge is
confining, it would be interesting to show explicitly that
these monopoles do form such a plasma. Indeed, since
the confinement is entirely driven by the largest cluster,
it is the monopoles in this cluster that should provide
a realisation of our simple picture in section~\ref{mon_conf}.
We should also be able to see in what way the monopoles 
belonging to the non-confining secondary clusters do not 
constitute such a plasma.

The fact that the string tension has a finite continuum limit, 
means that both the screening length and the monopole density
should also scale, i.e. that they should be constant
when expressed in physical units, up to lattice corrections 
that vanish as $a\to 0$. (The reader may be aware that
this is not what happens in the D=2+1 U(1) theory, but the 
peculiarities of that theory are not relevant to us here.)
In this subsection we shall study the scaling properties of 
the screening length and we shall do so separately first for 
all the monopoles, then for the largest cluster alone, and 
finally for the secondary clusters alone. (We have already shown 
that the monopole density of the largest cluster scales while 
that of the secondary clusters does not.) The more subtle
question of whether we really have a plasma rather
than, say, a distribution of dipoles, is something we
shall not touch upon here.

Before moving to the details of our calculations we
need to reconsider how our simple monopole plasma
picture might be modified in a realistic context. 
First an aside: we shall calculate the screening 
length in an approximation where we neglect the non-static
character of our monopoles. A more substantial point
is that there will exist excitations of the
lightest screening mass. Thus the magnetic flux
from a monopole will not decay as a simple exponential
in $r$. What we call the screening mass, $m_s$, will show
up in the asymptotic exponential decay, as $r\to\infty$,
of the flux:
\be
B(r) \propto \exp\{-{r\over{l_s}}\} ,
\label{B23}
\ee
where $l_s = 1/am_s$ is the screening length and $r$ is the 
distance from the monopole, all in lattice units.
At very large $r$, however, this flux is very small and
will have a negligible effect upon Wilson loops. So what 
is relevant to confinement is not this asymptotic
screening mass but rather the effective screening
mass that governs the decay of the flux at those distances 
where the flux is still sufficiently large to
efficiently disorder Wilson loops. This effective
screening mass will be some combination
of the lightest screening mass and its excitations. It
is this that we would like to see scale. 

Another complication will arise when we consider the 
screening properties of a subset of all the clusters.
Although we have assumed that these clusters are
independent, it is unlikely that this is really the case.
If we were in a U(1) field theory then a monopole would,
through the (dual) Coulomb interaction, affect other 
monopoles whether they belonged to the same cluster or not.
That is to say, all monopoles participate in the screening
of all other monopoles. If we focus on the screening of
the monopoles within some subset of clusters, and if only
the monopoles in that subset are allowed to participate
in the screening, then we will in general obtain an
incorrect screening length. And if the total fraction
of the monopoles that are excluded does not scale (as will be
the case, for example, when we consider only the largest 
cluster) then the extracted `screening length' might well
not scale either. This, as we shall shortly see, is 
what occurs in our case, despite the fact that we have no
reason to think that our U(1) fields are governed by a
simple U(1) effective action. 
Note that although the Coulomb interaction between
individual monopoles in different clusters might be 
important for screening, it is a weak high-order multipole 
interaction between well separated clusters. Since the
secondary clusters are compact objects ({\em vide} the
next section) this interaction should be weak
enough not to affect our derivation of eqn.~(\ref{A17}).

Suppose, then, that $B(r)$ is
the flux from a monopole. We expect that for large enough $r$
eqn.~(\ref{B23}) will hold. If we now plot
$- \ln B(r)$ against $r \arts$, then we expect to
see a linear rise at large $r$ whose slope is just
the inverse of the screening length in physical units, 
$\xi_s = l_s \arts$. Moreover if the the screening
length is constant in physical units, then this slope
should be independent of $\beta$. (All this assumes we
are in an infinite volume. In a finite periodic volume
one needs to make an obvious finite volume correction and
this we shall do.)
 
\begin{figure}
\begin{center}

\leavevmode
\epsfxsize = \figsize
\epsffile{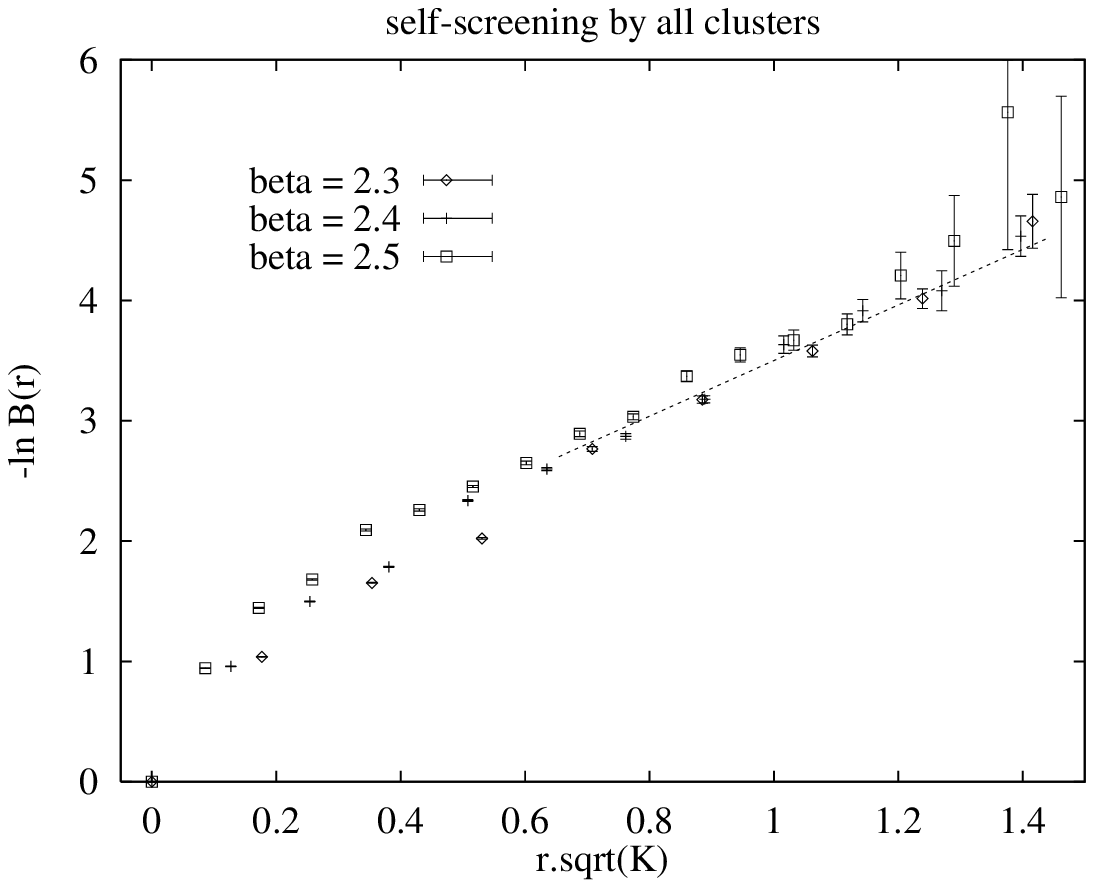}
\end{center}
\caption{Screening of flux by all monopoles as a function of distance 
in physical units for monopoles at $\beta = 2.3$, 2.4 and 2.5: with a 
linear fit shown.}
\label{fig_screening_all}
\end{figure}

In Figure~\ref{fig_screening_all} we produce such 
a plot using all the monopoles on the lattice. 
(Note that we normalise the monopole flux to unity.)
We see that indeed there is a linear rise
at large $r$, and that the slope is independent of $\beta$,
within statistical errors. That is to say, we find a scaling 
screening mass. If we now fit the combined data with a single 
scaling mass, we obtain $m_s \simeq 2.30~(10) \surd\sigma$.

As we can see in Figure~\ref{fig_screening_all} this
scaling screening mass only governs the decay of the
magnetic flux at large distances where the flux is small. 
If instead we look at the effective screening mass in
the range of $r$ where the flux is still large enough to 
disorder Wilson loops, say $\frac{1}{e} \le B(r) \le 1$, 
then we see that it does not scale. This should not be
a great surprise given that we have seen that the total
monopole density does not scale either, and that there is
a substantial number of monopoles, those in the secondary 
clusters, which do not appear to contribute to confinement. 

\begin{figure}
\begin{center}

\leavevmode
\hbox{%
\epsfxsize = 0.45\textwidth
\epsffile{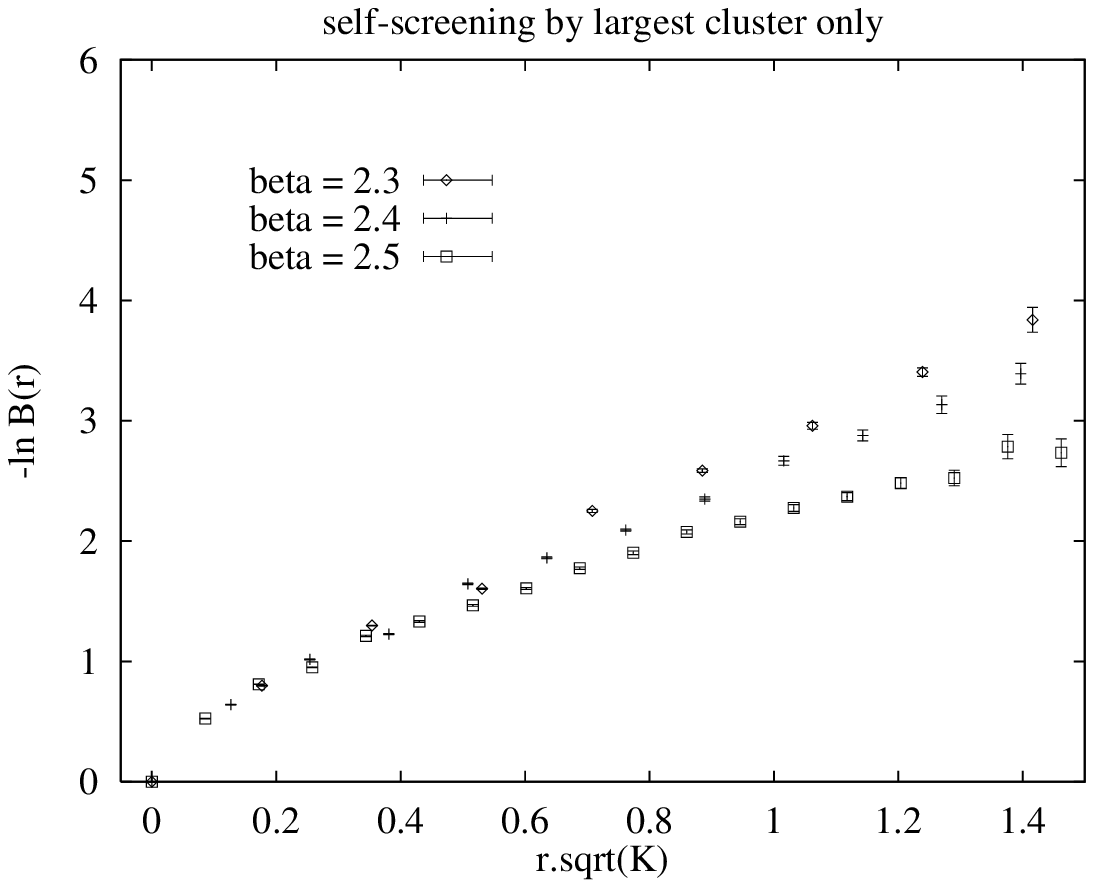}
\epsfxsize = 0.45\textwidth
\epsffile{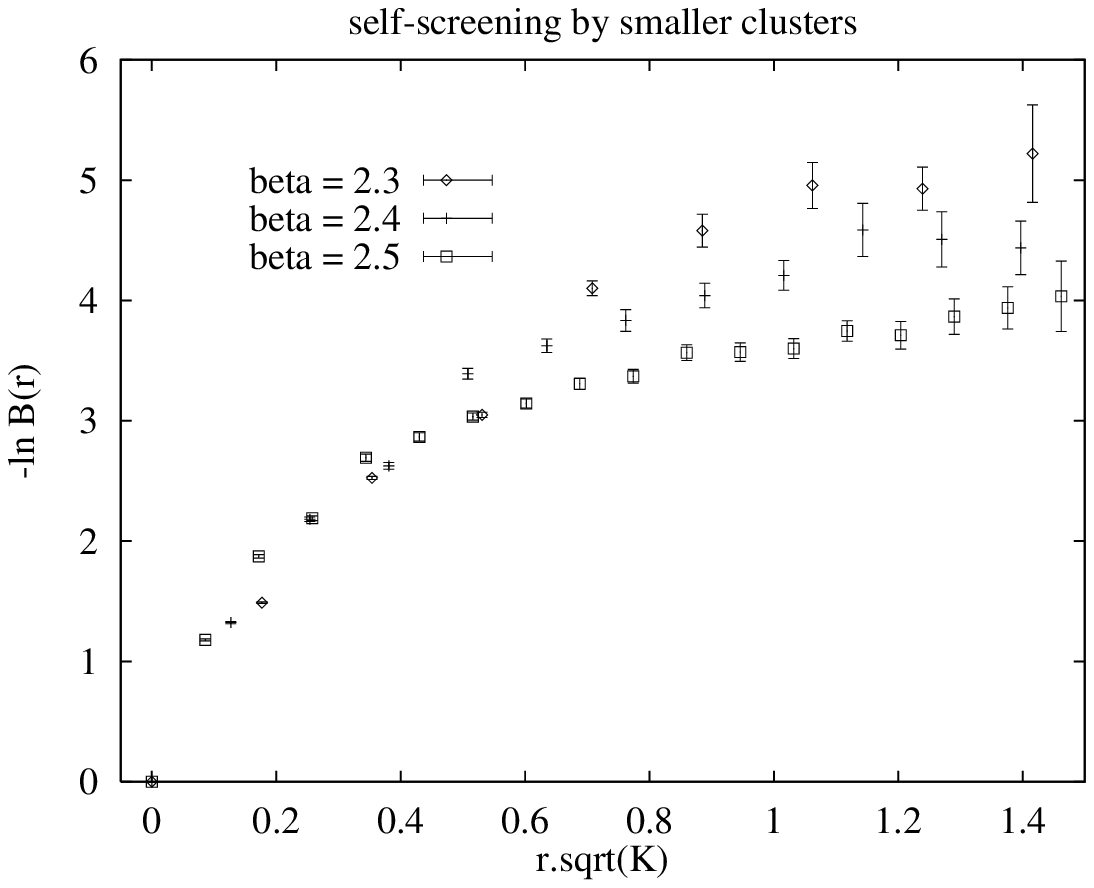}
}
\end{center}
\caption{Screening of flux as a function of distance in physical units for 
monopoles at $\beta = 2.3$, 2.4 and 2.5. The left hand plot uses monopoles 
from the largest cluster only; the right hand plot from the 
remaining, smaller clusters.}
\label{fig_screening_notall}
\end{figure}

Since we have found that it is the largest cluster that 
generates all of the string tension, and that the 
remaining clusters generate none of it, it is
interesting to repeat the analysis separately for these 
two subsets of the total monopole current. This we do in
Figure~\ref{fig_screening_notall}. We first note that in 
both cases the screening mass does not scale -- to the extent 
that one can identify a linear rise at large $r$.  Moreover the 
large-$r$ screening is much weaker than that obtained when we 
include all the clusters. This `under-screening' is what we would 
expect if there were Coulomb interactions between all the monopoles,
as discussed previously. For the secondary clusters the lack
of scaling persists down to the smaller values of $r$
which are relevant for confinement. For the largest cluster, 
on the other hand, the small-$r$ effective screening
masses do scale and we extract a value $m_s = 2.5 (1) \surd\sigma$, 
which is similar to the screening mass at large $r$ from all 
the clusters. This and the fact that the density of monopoles
in the largest cluster scales confirms that it is indeed the
monopoles in this cluster that provide the confining
monopole plasma.

\begin{figure}
\begin{center}

\leavevmode
\epsfxsize = \figsize
\epsffile{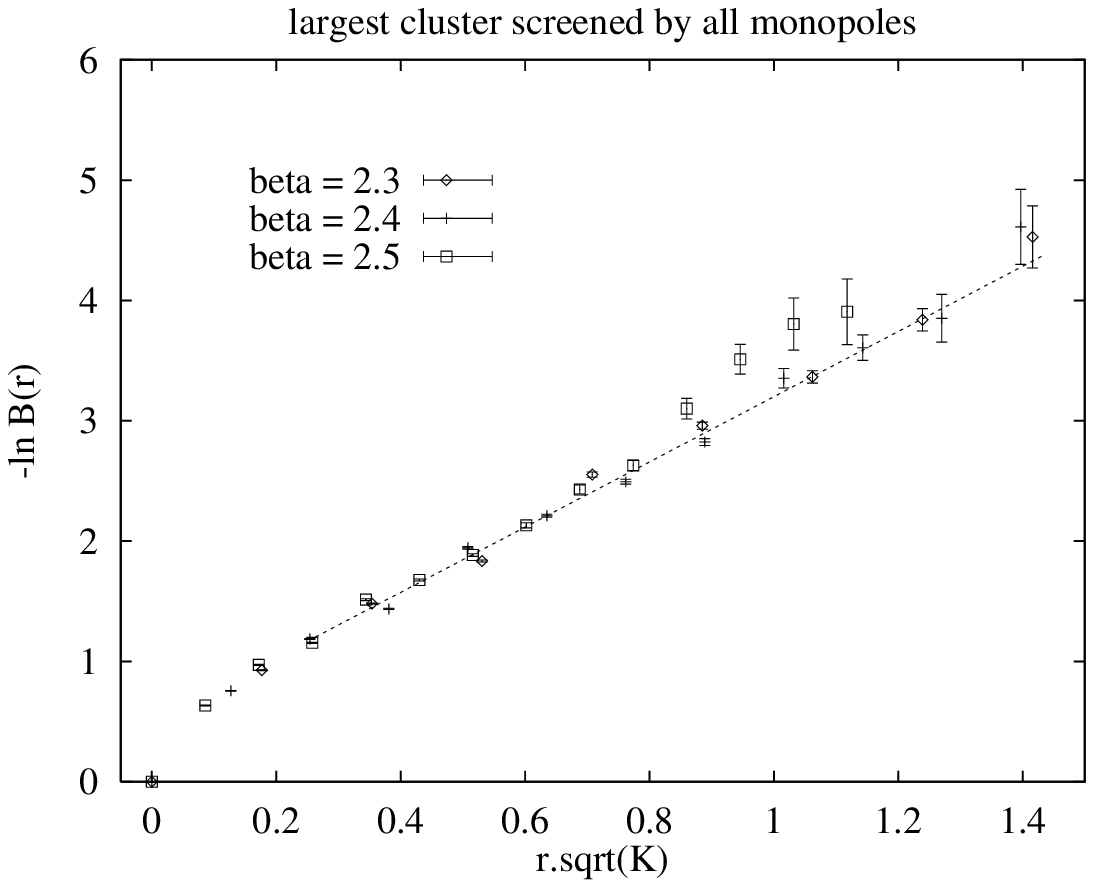}
\end{center}
\caption{Screening of flux from monopoles of the largest cluster by all 
other monopoles as a function of distance in physical units for 
monopoles at $\beta = 2.3$, 2.4 and 2.5; with a linear fit shown.}
\label{fig_screening_lg_all}
\end{figure}

Given the above discussion it would appear that the clearest
way to reveal the screening of the confining monopoles would
be to consider only the flux from those monopoles that are in the
largest cluster, but to include all monopoles in the screening
of that flux. This we do in Figure~\ref{fig_screening_lg_all}.
We now observe a flux that scales at all $r$. Moreover it can
almost be described by a single exponential at all $r$. 
(We should use a lattice version of the Coulomb interaction
at very small $r$, but we ignore this potential improvement
here.) We extract a screening mass of $m_s = 2.71 (15) \surd\sigma$,
which is roughly consistent with our other values.

Before leaving this topic, it is interesting to ask if
this screening mass has anything to do with the spectrum
of the underlying $SU(2)$ theory. Abelian dominance
suggests that this is just the effective gluon mass in the
maximally Abelian gauge. There have been speculations in the
past that gauge-fixed quark mass calculations (typically
performed in the Landau gauge) are telling us about
the constituent quark mass. So the analogous speculation here
would be that our screening mass is related to a constituent
gluon mass. It is therefore amusing to note that the lightest 
glueballs in the SU(2) theory are the scalar and the tensor, 
with continuum masses of $m_{0^{+}} = 3.87(12) \  \surd\sigma$
and $m_{2^{+}} = 5.63(11) \  \surd\sigma$ respectively
\cite{michael88,gluesu2}.
In a simple constituent gluon picture of the low-lying
glueball spectrum one would expect these states to
arise from two gluons in an $L=0$ state, with the spins
aligned to give the tensor and anti-aligned to give the
scalar. Thus to leading order in the spin-spin interaction
the scalar and tensor masses will be equally split from
the mass that they would 
possess if the spin-spin interaction were not present.
The observed splitting from the average value is 
$\sim \pm 20\%$ which is small enough for the leading order
argument to be plausible. The average mass value will then
equal that of two constituent gluons, neglecting binding
energies (which have to be small if a constituent gluon
picture is to have any chance of making sense). We observe 
that our screening mass is indeed in the right ball-park 
to be thought of as such a `constituent gluon' mass.

\vskip 0.25in

To sum up this section, we have established that the largest cluster
is a quite different animal from all the other clusters. It permeates
the entire volume, has a constant density and screening length in
physical units, and drives confinement. It would seem natural to think
of this largest cluster as being a simple example of na\"{\i}ve
percolation at work. But, as we have seen, this is not the case. If
percolation is at work, it is at work within the SU(2) field
configurations of which our monopoles are but a skeletal
representation.

\section{The smaller clusters}
\label{sec_smaller}

\subsection{The cluster spectrum}

\begin{figure}
\begin{center}

\leavevmode

\noindent
\epsfxsize = 0.32\textwidth
\epsffile{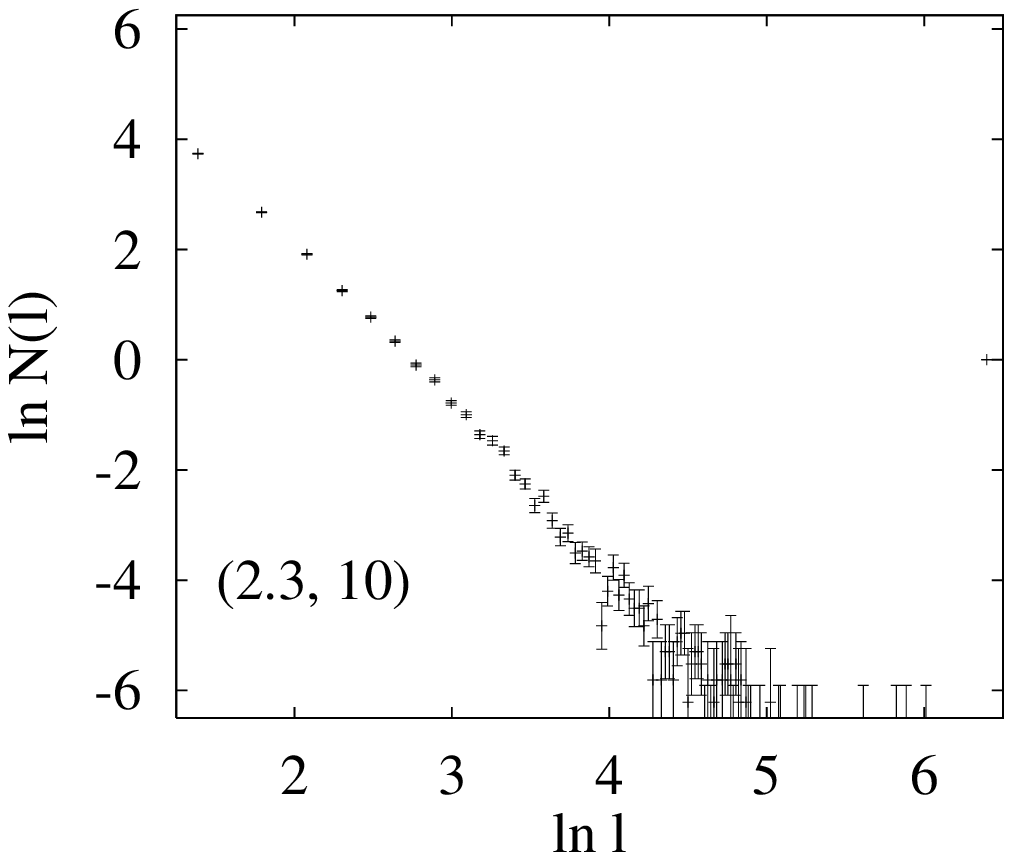}
\epsfxsize = 0.32\textwidth
\epsffile{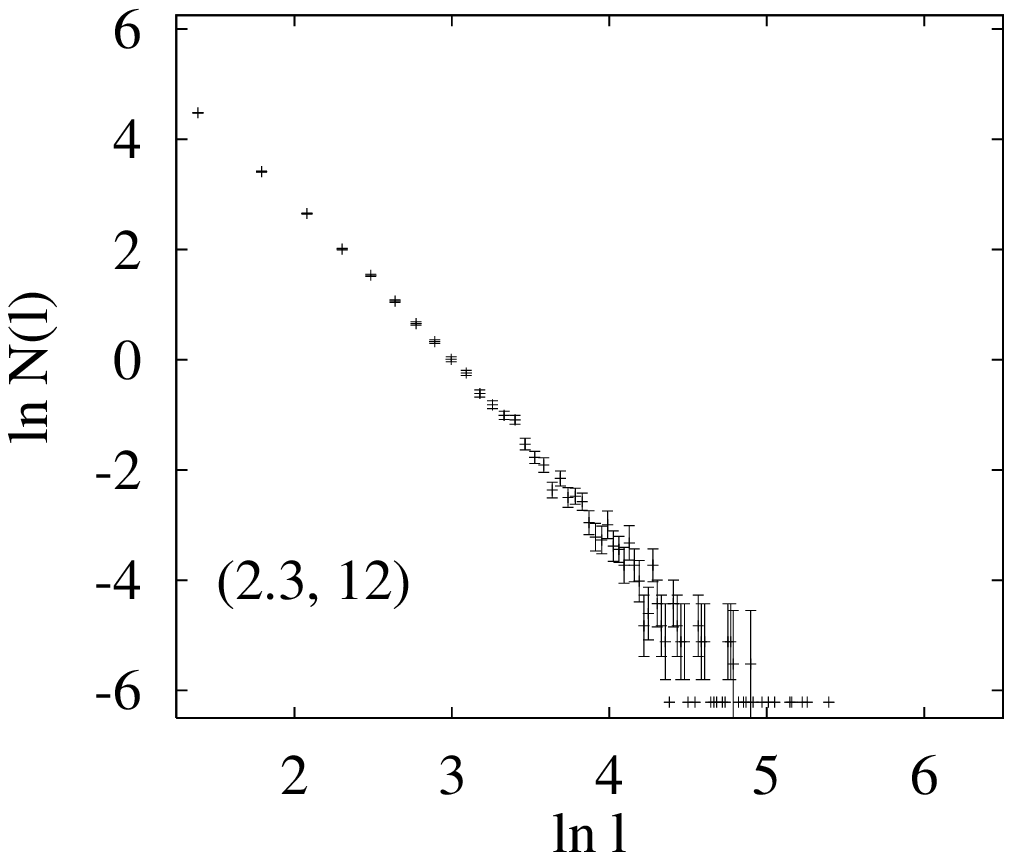}
\epsfxsize = 0.32\textwidth
\epsffile{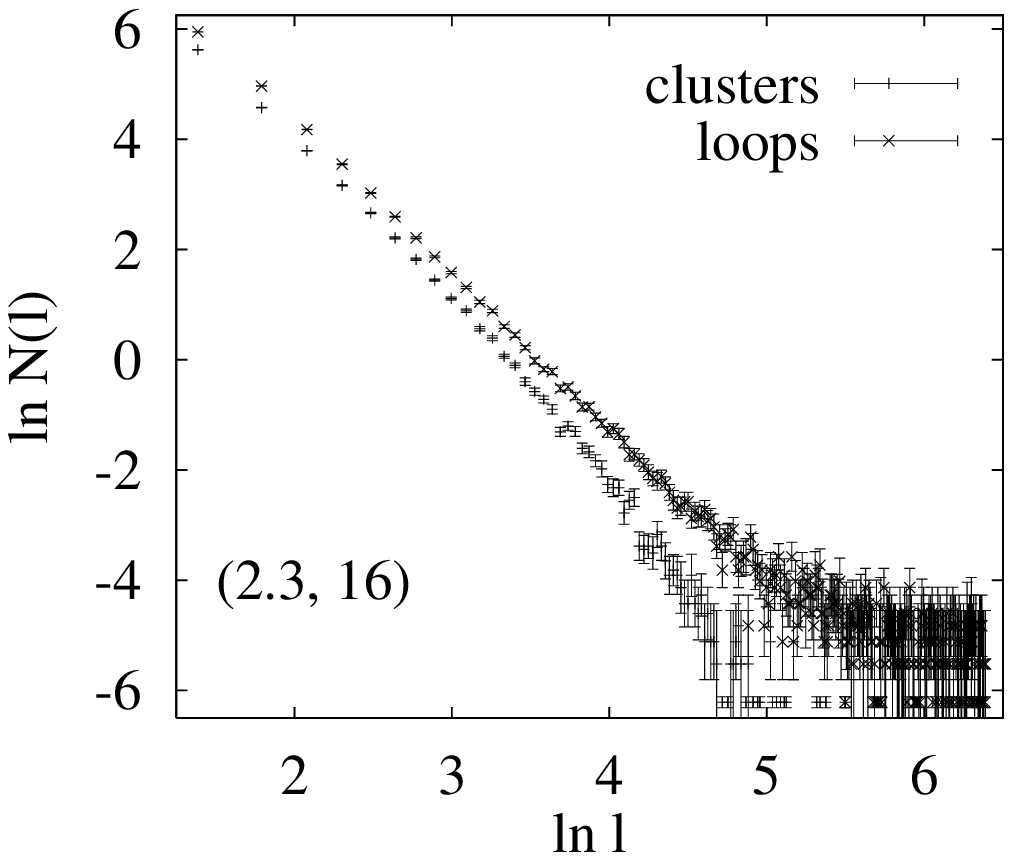}
\end{center}
\caption{Monopole cluster spectra at $\beta = 2.3$ on $L=10$, 12 and 16.
The loop spectrum is shown for comparison on $L=16$.}
\label{fig_spectrum_volume}
\end{figure}

\begin{figure}
\begin{center}

\leavevmode

\noindent
\epsfxsize = 0.45\textwidth
\epsffile{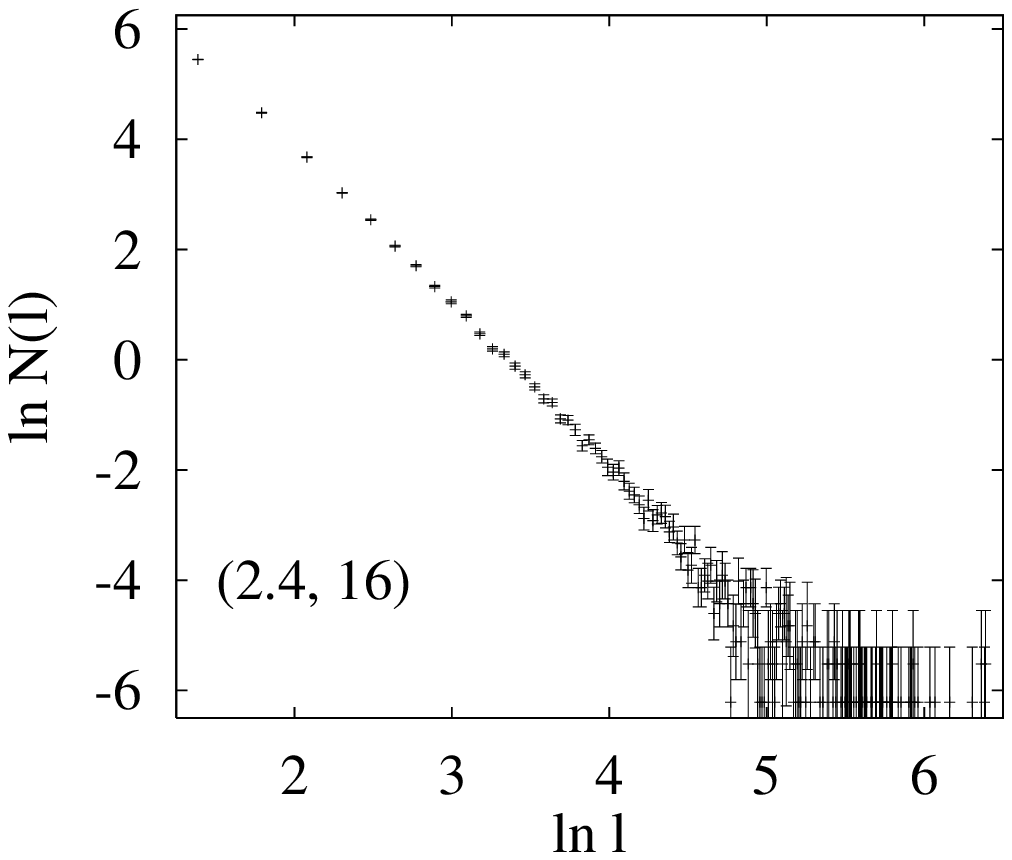}
\epsfxsize = 0.45\textwidth
\epsffile{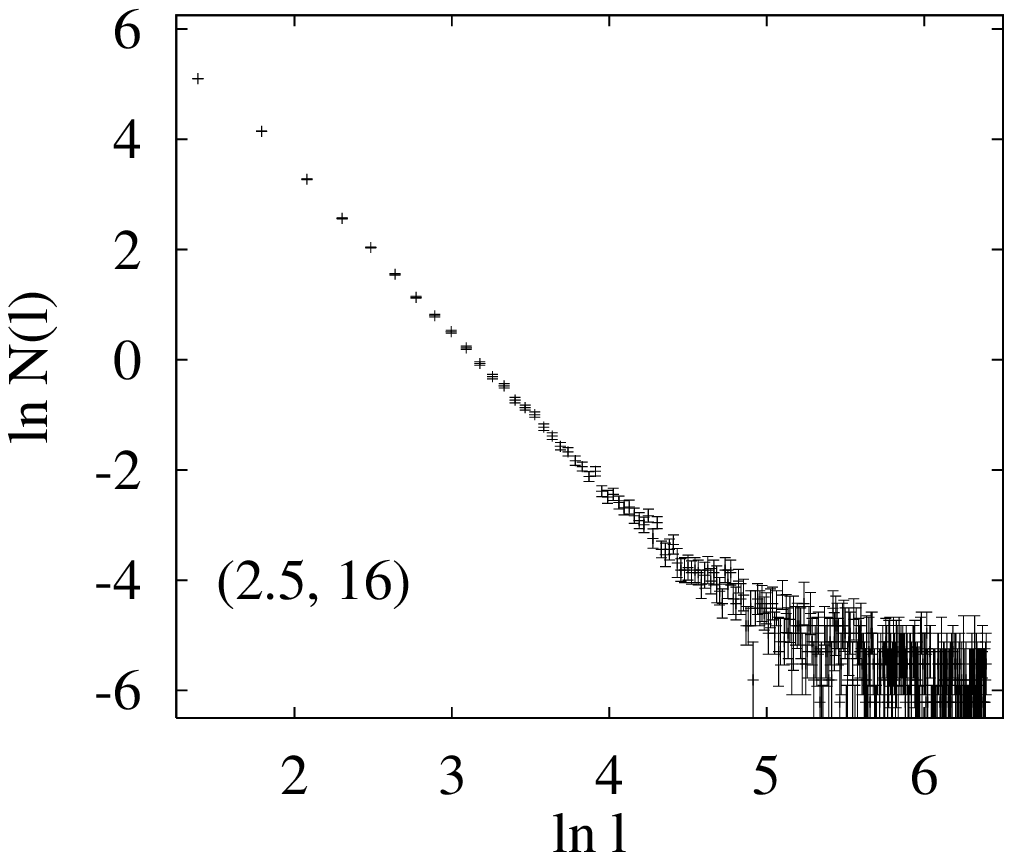}
\end{center}
\caption{Monopole cluster spectra on $L=16$ at $\beta = 2.4$ and 2.5.
The equivalent for $\beta = 2.3$ is given in 
Figure~\ref{fig_spectrum_volume}.}
\label{fig_spectrum_spacing}
\end{figure}

We have frequently stated that the number of secondary clusters falls
off approximately as a cubic power of the cluster length $l$. Some
evidence for this is shown in Figure~\ref{fig_spectrum_volume} where
we display the spectra for three different volumes at
$\beta=2.3$. These spectra all fall roughly as $1/l^3$ for the range
of $l$ where our calculations are accurate. There is also some
evidence that at very large $l$ there is a change in the functional
form.  There seem to be finite size effects there, and the indication
is that on large enough volumes, the spectrum for very large values of
$l$ might fall off more steeply. More accurate calculations than ours are
needed to determine whether this is indeed so.  In
Figures~\ref{fig_spectrum_volume} and~\ref{fig_spectrum_spacing} we
also display the spectra obtained on $16^4$ lattices at $\beta=2.3$,
2.4, and 2.5.  This shows that the $\sim 1/l^3$ behaviour does not depend
on $a$. We note that as $a$ decreases, the very large $l$ end of the
spectrum on the $16^4$ lattices appears to show finite volume effects;
perhaps not surprising since $16a(\beta=2.5) \simeq 8a(\beta=2.3)$.
The slight curvature of the spectrum leads to the fit parameters of
the power law depending weakly on the range of $l$ that we choose to
fit. Nevertheless, we are able to conclude that all the fits to our
data have an exponent in the range $\gamma \in [2.85,3.15]$.

In Figure~\ref{fig_spectrum_volume} we also show for comparison the
spectrum obtained when the monopole currents are divided into
loops. The main difference is in the normalisation; there are more
loops of a given size than clusters. Some proportion of the small
loops of a given size will be part of larger clusters, and in
particular the largest cluster on the lattice. It is interesting,
nonetheless, to note that the exponent of the power law for the loop
spectrum is in general slightly smaller than that for the more
fundamental (we believe) cluster spectrum.

The simplest way to understand this cluster spectrum would be if
there were, in essence, only the one current cluster in each field
configuration (the very large cluster that we described in the
previous section) and that the secondary clusters then arose when
small portions of this largest cluster were randomly `pinched'
off. The power law spectrum would then have to arise from the relative
probability of pinching off portions of the largest cluster of
different lengths. Were this the case, the number of clusters of a
given length on a configuration (particularly the smallest and most
numerous) would be expected to be proportional to the (remaining)
length of the largest cluster from which they were
formed. Unfortunately this turns out not to be even approximately the
case, there being no correlation, either positive or negative.
(We might also expect the smaller clusters to be preferentially
located near current links of the largest cluster, although we did not
test this.)

\subsection{Cluster sizes}

\begin{figure}
\begin{center}

\leavevmode
\epsfxsize = \figsize
\epsffile{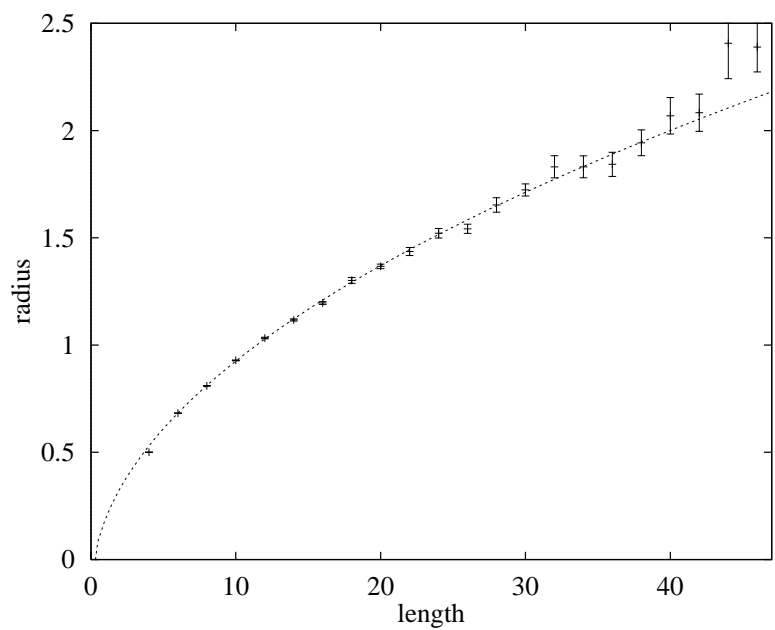}

\end{center}
\caption{The cluster radii and a fit $r_{\eff} = s + t \surd l$
at $\beta = 2.3$ on $L = 12$.}
\label{fig_ext_plus_fit}
\end{figure}

What do we know about the sizes of these secondary clusters?
We can estimate the cluster radius
using the first moment of the current links about the centroid
of the cluster. If the cluster were composed of $n$ current
links of charge $\{ j^i : i = 1,n \}$ with centres at $\{
x_\mu^i \}$, then the centroid is
\be
{\bar{x}}_\mu = \frac{1}{l} \sum_{i=1}^n x_\mu^i \left| j^i \right|
\label{A24}
\ee
where the length is
\be
l = \sum_{i=1}^n \left| j^i \right|.
\label{A25}
\ee
The distance of the centre of a link from the centroid is
$d^i$, and the effective radius of the cluster is
\be
r_{\eff} = \frac{1}{l} \sum_{i=1}^n d^i \left| j^i \right|.
\label{A26}
\ee
We plot this as a function of length in
Figure~\ref{fig_ext_plus_fit}, and find that it is well fitted
by the functional form 
	$r_{\eff}(l) = s + t \surd l$.
This suggests that the monopole is essentially performing
a random walk. Is the step size of this walk fixed in 
lattice or in physical units? If it were fixed in lattice units
we would expect $t$ to be independent of $a$. 
If the step size were fixed in physical units then we would
have $r \arts \propto \sqrt{l \arts }$, and so
would expect $t^2 \propto 1 / \arts$. Our calculations, examples
of which are presented in
Table~\ref{tab_sqrt_fits}, show us that the coefficient $t$ varies
very weakly with $a$ if at all. There is some variation in our
fitted value of $t$ depending on the range of $l$ used.  But our
overall conclusion is that if we insist on parameterising $t$ by
a power of $a$ then that power is small: 
$t^2 \propto (\arts)^{-{1\over 8} \pm {1\over 8}}$ 
So, although there is some room for a residual weak dependence on $a$, 
the evidence is that the step size in the cluster random walk does 
not know about physical units. 

We note that the values of $l$ where we saw, in
Figures~\ref{fig_spectrum_volume} and~\ref{fig_spectrum_spacing}, 
evidence of finite size effects in $N(l)$, 
do indeed correspond to cluster
sizes, $r_{\eff}$, that might plausibly feel the 
boundaries of our periodic lattices. 

\subsection{Scaling properties}

We now turn to the normalisation of the spectrum of these secondary
clusters and ask what scaling properties it possesses.  We have
already seen that the total density does not scale: that is, the total
length of the secondary clusters is not proportional to the volume
when both are expressed in physical units.  This in itself is no
surprise, however.  When we decrease $a$ by a factor of say 2, then
the total current length acquires an additional contribution that is
$\sim \int_4^8 l N(l)dl$ in units of the smaller lattice
spacing. (Since the smallest cluster has length $4a$.)
This will be a significant contribution because the spectrum grows
rapidly at small $l$. So if nothing else, we expect a significant
scaling violation from the growing tail of ultraviolet clusters and
any test of scaling must take this into account. The simplest form of
physical scaling would be to consider only those clusters whose length
is larger than some fixed physical length $l_p$, i.e.  $l \ge
l_p / \arts$, and then to demand that the total length of these
clusters is proportional to the volume when both are expressed in
physical units. We now see what this implies for the observed spectrum
\be
N(l) = \frac{C(L,a)}{l^\gamma}.
\label{A27}
\ee
with $\gamma \sim 3$. (For these purposes any deviation at
very large $l$ is negligible, and the deviations at
the small $l$ ultraviolet scale are irrelevant.)
Scaling would imply
\be
\arts \int_{{l_p}\over{\arts}}l N(l)dl = 
\arts \int_{{l_p}\over{\arts}} l \frac{C(L,a)}{l^\gamma}dl
\propto {(L \arts)}^4
\label{28}
\ee
which requires
\be
C(L,a) \propto L^4 (\arts)^{5-\gamma}.
\label{29}
\ee
This is to be contrasted with what we  should expect
if these secondary clusters only knew about the ultraviolet length
scale, $a$: $C(L,a) \propto L^4$. As we have already seen, in
Table~\ref{tab_lengths},
the factor of $L^4$ is certainly there. What is at issue is
the dependence on $\arts$. Scaling requires that
the quantity
\be
c_p^1 = \frac{C}{L^4.(\arts)^{5-\gamma}}
\label{30}
\ee
should be independent of $\beta$. 
In Table~\ref{tab_pow_law} we show the values of $c_p^1$ that we have
obtained on our $L=16$ lattices both when we use the value of $\gamma$
obtained form the power law fit, and when we impose a fixed value
$\gamma = 3$ at all values of $\beta$ (as our above analysis assumes).
As we see, rather than being constant $c_p^1$ increases roughly as
$1 / \artsq$. This is what one expects (with $\gamma \simeq 3$)
if the clusters know only of the ultraviolet scale.

We have tested a particular formulation of scaling which,
na\"{\i}vely, would seem to be the most reasonable. It is not unique,
however. A plausible alternative would be to focus on the total number
of clusters instead of their total length. If we consider the total
number of clusters whose length is greater than some constant in
physical units, then in fact we find the same criterion as
above. To get something different we might, for example, ask (as in
\cite{hart97a})
whether perhaps it is $N(l)$ itself that scales with the physical
volume, when $l$ is chosen fixed in physical units.  This would
require $c_p^2 = \frac{C}{L^4.(\arts)^{4-\gamma}}$ to be independent
of $\beta$.  In Table~\ref{tab_pow_law} we show this, again for
$\gamma$ from the power law fits at different $\beta$ and with a
single, imposed value of $\gamma = 3$. Using the fitted $\gamma$, this
quantity appears to scale much better. This result is not robust;
imposing a fixed $\gamma$, however, where the statistical errors are
less, this scaling appears less good. Without some argument for
keeping the measure $dl$ in lattice units (which is what we have just
done), however, it does not really make sense as a scaling
criterion. It seems that if we take the secondary clusters at face
value, they certainly do not have the right scaling properties to be
physical objects.

Given that the secondary clusters do not scale as `physical objects',
we can ask whether they scale as purely lattice artifacts. If so we
would expect the total current length to be $\propto L^4$ but to be
independent of $a$. So if we focus on the $L=16$ lattices in
Table~\ref{tab_lengths}, we would expect $(l_{\tot}-l_{\max})$ to be
independent of $\beta$.  In fact the values are 3226, 2997, and 2244
at $\beta=2.3$,~2.4 and~2.5 respectively. We know that the $\beta=2.5$
value is suppressed below its true value because there is some overlap
between the largest cluster and the secondary spectrum: so $l_{max}$
is certainly overestimated. Nonetheless, even allowing for that, there
does appear to be some significant $a$ dependence, $C(L,a) \propto L^4
(\arts)^{0.2-1.0}$, but it is quite weak suggesting that the spectrum
is influenced more by the ultraviolet than by the physical length
scale.

\subsection{Clusters as 4-balls}
 
Of course the monopole currents are only images, through gauge-fixing,
of some unknown structures in the SU(2) gauge fields. It is the latter
that one would hope to be physical. In fact our observed cluster
spectrum does provide some intriguing hints as to what these
structures might be. As we have seen, a monopole cluster of length $l$
is localised within a region in space-time of size $r \simeq t \surd
l$. We call such an object a `4-ball' for obvious reasons. What is the
spectrum, $N_B(r) dr$, of these 4-balls? It is easy to see that if the
radius is related to the length by $r \propto \surd l$, the cluster
spectrum $ N(l) dl = C/l^3 $ translates into the following 4-ball
spectrum:
\be
N_B(r) dr = C_B \frac{dr}{r} \times \frac{1}{r^4}.
\label{31}
\ee
We recognise this to be simply the general scale invariant
distribution of objects of radius $r$ in four dimensions. (Such an
object takes up a volume $\sim r^4$ and hence there are $\sim 1/r^4$
ways of placing it in a unit volume. And $dr/r$ is a scale-invariant
measure.) This is precisely the formula one has for the density of
instantons, before one includes the effects of the scale anomaly
through the running of the coupling. Since we know that an isolated
instanton, when projected to the maximally Abelian gauge, generates a
monopole current loop of size comparable to its core size
\cite{hart96},
it would be tempting to put forward the elegant hypothesis that these
4-balls are just SU(2) instantons and that the secondary clusters are
simply the associated monopole loops.  Unfortunately things cannot be
so simple. Although we do not know how the scale-breaking affects the
instanton density at large sizes, we do know what it does to the
distribution at small sizes: the $dr/r^5$ is transformed into
$r^{7/3}dr$. This is nothing at all like our 4-ball number density.

It is interesting to repeat our previous scaling analysis, but this
time assuming that it is the 4-balls that are physical rather than the
clusters themselves. That is to say, we impose that the number of
4-balls of radius larger than some fixed length in physical units,
should be proportional to the physical volume. It is easy to see that
this implies that $C_B(L,a) \propto L^4$. One obtains the same result,
however, if one constrains the density to be constant in lattice
units, or any other units, because the 4-ball density just reflects
na\"{\i}ve dimensional counting. Thus we expect rather generally that
\be
N(l) dl =  \frac{C(L,a) dl}{l^3}
\propto \frac{L^4}{t^4} \frac{dl}{l^3}
\label{A32}
\ee
where we have gone from the 4-ball density to the cluster
spectrum using $r \sim t\surd l$. This implies that
\be
C(L,a) \propto \frac{L^4}{t^4} 
\propto L^4 (\arts)^{0.25 \pm 0.25}
\label{A33}
\ee
using our results for the $a$-dependence of $t$. 
This weak $a$-dependence is entirely consistent with
what we observe for the cluster spectrum:
$C(L,a) \propto L^4 (\arts)^{0.2 \to 1.0}$
Thus the spectrum of secondary clusters is consistent, 
in every respect, with arising from a scale-invariant 
density of 4-balls.

As an aside, we note that the largest cluster from the
distribution $N(l)$ has a length 
$l_{\sec} \propto C(L,a)^{1\over{\gamma -1}}$.
Putting in $\gamma=3$ and the form for $C(L,a)$
as in the previous paragraph, we see that
$l_{\sec} \propto L^2 (\arts)^{0.1 \to 0.25}$.
By contrast the length of the largest cluster varies as:
$l_{\max} \propto L^4 (\arts)^3$. From this we
see that if we wish to maintain $l_{\max} \gg l_{\sec}$
as $a \to 0$ then the lattice size in physical units
must grow roughly as 
$(1 / \arts )^{1/2}$. 
Thus, for example,
the two types of clusters begin to overlap on our
$16^4$ lattice at $\beta=2.5$ (rendering some of the
calculations there ambiguous) despite the fact that
they did not do so on the  $8^4$ lattice at $\beta=2.3$.
This is something we did not, of course, anticipate
when originally choosing our lattice sizes.

\section{Summary}
\label{sec_summary}

In this paper we have shown that the magnetic monopole
currents that we obtain, when gauge fixing SU(2) fields to the
maximally Abelian gauge, divide into two quite distinct classes
(on large enough volumes): 
a single very large cluster and a distribution of very much smaller 
clusters.

The very large cluster has a length that is proportional to the
space-time volume when both are expressed in physical units.  Moreover
we have shown that it is this cluster that generates the string
tension. 
We have also seen that, within this largest cluster, the 
effective screening length relevant to confinement
is constant in physical units.
Thus it is this cluster that represents all the
interesting infrared physics of the SU(2) fields.

That there is always just one very large cluster is 
a significant fact since, as we saw, 
even under a minimal amount of smoothing this cluster
readily hives off secondary clusters that are much larger than
those that we observe in the unsmoothed fields.  

The secondary clusters are localised compact objects obtained 
by the monopole performing a random walk on the length of 
the lattice spacing. This is in contrast to the largest cluster
whose observed scaling demands that the step size be on the
length of the physical length scale. These secondary clusters
contribute nothing to the string tension even where they
constitute a sizeable fraction of the total magnetic current.

One might be tempted to ignore these secondary clusters as being of no
physical importance. They do seem quite remarkable in at least one
respect, however. When one treats them as localised objects in
space-time (`4-balls'), one finds that the number density is of the
simplest scale-invariant form. This is reminiscent of classical
instantons, but unfortunately incompatible with the real instanton
density at small distances.

The calculations of this paper can be improved upon in many
ways. In particular better calculations could clarify what 
happens to the distribution of secondary clusters at very large
$l$ and it would be useful to calculate the 4-ball number
density directly (as we would have done if we had not deduced
their relevance after completing the simulations). 

The monopole content of the vacuum thus seems to split up into two
types of cluster. First there is the confining cluster that knows
about the physical length scale (ultimately due to the breaking of
scale invariance) but does not seem to know anything at all about the
lattice length scale.  Secondly there are the other, smaller
clusters. These can be thought of as compact objects that satisfy a
scale invariant distribution: while they know about the lattice
spacing, they apparently know little about the breaking of scale
invariance. This is unexpected and puzzling, because these clusters
should somehow reflect fluctuations in the SU(2) fields. Of course,
because the gauge-fixing procedure is completely non-local, it is
possible that the monopoles we observe only reflect an effective
theory that possesses the same infrared physics as the non-Abelian
theory. To resolve this puzzle would be of interest.

\vspace{0.20in}
\noindent {\bf Acknowledgements}

The work of A.H. was supported in part by United States Department of
Energy grant DE-FG05-91 ER 40617. M.T. has been supported by United
Kingdom PPARC grants GR/K55752 and GR/K95338.

\newpage

\begin{table}
\begin{center}
\begin{tabular}
{r 
r@{.}l@{\hspace{0.5\tabcolsep}(}r@{)\hspace{2\tabcolsep}}
r@{.}l@{\hspace{0.5\tabcolsep}(}r@{)\hspace{2\tabcolsep}}
r@{.}l@{\hspace{0.5\tabcolsep}(}r@{)\hspace{2\tabcolsep}}
}
\hline \hline
$ r $ &
  \multicolumn{3}{c}{$\sigma_{\uone}(r)$} &		
  \multicolumn{3}{c}{$\sigma_{\mon}(r)$} &		
  \multicolumn{3}{c}{$\sigma_{\diff}(r)$}\\		
\hline
2 &  0&1561 &  18 & 0&0673 &  5 & 0&0894 & 14 \\
3 &  0&1103 &  28 & 0&0651 &  6 & 0&0348 & 20 \\
4 &  0&0983 & 132 & 0&0649 & 12 & 0&0132 & 41 \\
5 &  0&0259 & 354 & 0&0628 & 21 & 0&0040 & 80 \\
6 &  \multicolumn{3}{c}{} & 0&0621 & 37 & -0&0056 & 172 \\
\hline \hline
\end{tabular}
\end{center}
\caption{The effective string tension obtained from Creutz ratios of
size $r$; as obtained from the U(1) fields, the monopole
clusters and from the difference of the U(1) and monopole
fluxes. 
All are obtained from an ensemble of 500 configurations on $16^4$ 
lattices at $\beta=2.4$}
\label{tab_sigma_diff}
\end{table}

\begin{table}
\begin{center}
\begin{tabular}
{rr 
r@{\hspace{0.5\tabcolsep}(}r@{)\hspace{2\tabcolsep}}
r@{.}l@{\hspace{0.5\tabcolsep}(}r@{)\hspace{2\tabcolsep}}
r@{\hspace{0.5\tabcolsep}(}r@{)\hspace{2\tabcolsep}}
r@{.}l@{\hspace{0.5\tabcolsep}(}r@{)\hspace{2\tabcolsep}}
r@{\hspace{0.5\tabcolsep}(}r@{)\hspace{2\tabcolsep}}
}
\hline \hline
$\beta$ & $L$ & 
  \multicolumn{2}{c}{$l_{\tot}$} &		
  \multicolumn{3}{c}{$l_{\tot} / L^4$} &		
  \multicolumn{2}{c}{$l_{\max}$} &		
  \multicolumn{3}{c}{$l_{\max} / L^4$} &		
  \multicolumn{2}{c}{$l_{\sec}$} \\	
\hline
2.3 & 8  & 
	840 & 4 & 0&205 & 1 & 624 & 5  &    0&152 & 11 & 35 & 10 \\
 & 10 &
	2054 & 5 & 0&205 & 1 & 1558 & 58 &   0&156 & 6 & 42 & 11 \\
 & 12 &
	4230 & 7 & 0&204 & 1 & 3200 & 110 &   0&154 & 5 &  49 & 10 \\
 & 16 &
	13394 & 18 & 0&204 & 1 & 10168 & 141 & 0&155 & 2 &  67 & 9 \\
\hline
2.4 & 10 &
	1100 &  5 & 0&110 & 1 &  584 &  70 & 0&058 & 7 & 83 & 26 \\
    & 12 &
	2288 & 12 & 0&110 & 1 & 1277 & 104 & 0&062 & 5 & 116 & 39 \\
    & 14 &
	4228 & 10 & 0&110 & 1 & 2441 & 141 & 0&064 & 4 & 121 & 48 \\
    & 16 &
	7184 & 18 & 0&110 & 1 & 4187 & 177 & 0&064 & 3 & 125 & 38 \\
\hline
2.5 & 16 &
	3583 & 16 & 0&055 & 1 &	1339 & 123 & 0&020 & 2 & 255 & 40 \\
\hline \hline
\end{tabular}
\end{center}
\caption{The total length of the current, $l_{\tot}$, and 
its scaling with the lattice volume. Ditto for the largest 
cluster, $l_{\max}$. The length of the second 
largest cluster is also listed.}
\label{tab_lengths}
\end{table}

\begin{table}
\begin{center}
\begin{tabular}
{l r@{.}l@{\hspace{0.5\tabcolsep}(}r@{)\hspace{2\tabcolsep}}
r@{.}l@{\hspace{0.5\tabcolsep}(}r@{)(}r@{)\hspace{2\tabcolsep}}
r@{.}l@{\hspace{0.5\tabcolsep}(}r@{)(}r@{)\hspace{2\tabcolsep}}
r@{.}l@{\hspace{0.5\tabcolsep}(}r@{)\hspace{2\tabcolsep}} }
\hline \hline
$L = 16$: & 
  \multicolumn{3}{c}{$a^2 \sigma_{\tot}$} &		
  \multicolumn{4}{c}{$a^2 \sigma_{\max}$} &		
  \multicolumn{4}{c}{$a^2 \sigma_{\hbox{\tiny rest}}$} &		
  \multicolumn{3}{c}{$P$} \\		
\hline
$\beta = 2.3$: & 0&128 & 5 & 0&124 & 3 & 0 & 0&000 & 1 & 0 & 0&75 & 1 \\ 
$\beta = 2.4$: & 0&067 & 2 & 0&058 & 2 & 0 & 0&001 & 1 & 0 & 0&57 & 1 \\
$\beta = 2.5$: & 0&034 & 2 & 0&024 & 2 & 2 & $<0$&005 &{-} & 2 & 0&37 & 1 \\
\hline \hline
\end{tabular}
\end{center}
\caption{The monopole string tensions from all the clusters, the 
largest and from the remainder. The second error is the systematic bias 
from correcting the winding number. $P$ is the percolation parameter.}
\label{tab_sigma_prune}
\end{table}

\begin{table}
\begin{center}
{\small
\begin{tabular}
{r
*{3}{r@{.}l@{\hspace{0.5\tabcolsep}(}r@{)\hspace{2\tabcolsep}}}
*{2}{r@{\hspace{0.5\tabcolsep}(}r@{)\hspace{2\tabcolsep}}}
r@{.}l@{\hspace{0.5\tabcolsep}(}r@{)\hspace{2\tabcolsep}}
r@{\hspace{0.5\tabcolsep}(}r@{)\hspace{2\tabcolsep}}
r@{.}l@{\hspace{0.5\tabcolsep}(}r@{)\hspace{2\tabcolsep}}
}
\multicolumn{22}{c}{$\beta = 2.3$, $L = 12$} \\
\hline \hline
\multicolumn{1}{c}{$s$} & 
\multicolumn{3}{l}{$a^2\sigma$} &	
\multicolumn{3}{l}{\% curr.} &
\multicolumn{3}{l}{$P$} &
\multicolumn{2}{l}{$l_{\max}$} &
\multicolumn{2}{l}{$l_{\sec}$} &
\multicolumn{3}{l}{$n_C$} &
\multicolumn{2}{l}{$l^\prime_{\sec}$}  &
\multicolumn{3}{l}{$n^\prime_C$} \\
\hline
0 & 	0&128$^{\dagger}$ & 5 &	100&0 & 0 &	75&7 &  3 &	
	3200 & 22 & 	49 & 10 &	156&9 & 7 &
			49 & 10 &	156&9 & 7 \\	
1 & 	0&124 & 6 &	 56&7 & 2 &	78&5 &  4 &
	1882 & 65 &	85 & 32 &	56&5 &  3 &
			29 & 3  &	36&2 &  3 \\
2 &	0&122 & 5 &	 40&7 & 1 &	76&2 &  8 &
	1314 & 94 &	141 & 48 &	27&3 &  2 &
			 22 & 3 &	13&6 &  2 \\
3 &	0&124 & 5 &	 32&6 & 1 &	74&0 &  8 &
	1021 & 74 &	163 & 44 &	16&3 &  2 &
			18 & 2 &	 6&8 &  1 \\
5 &	\multicolumn{3}{c}{} &	 24&2 & 1 &	70&8 &  9 &
	725 & 64 &	160 & 39 &	8&9 &   2 &
			11 & 2 &	2&7 &   1 \\
10 & 	\multicolumn{3}{c}{} &	 15&5 & 1 &	68&4 & 11 &
	449 & 50 &	125 & 27 &	4&4 &   1 &
			  3 & 2 &	1&2 &   1 \\
\hline \hline
\end{tabular}

\vspace{2ex}
\begin{tabular}
{r
*{3}{r@{.}l@{\hspace{0.5\tabcolsep}(}r@{)\hspace{2\tabcolsep}}}
*{2}{r@{\hspace{0.5\tabcolsep}(}r@{)\hspace{2\tabcolsep}}}
r@{.}l@{\hspace{0.5\tabcolsep}(}r@{)\hspace{2\tabcolsep}}
r@{\hspace{0.5\tabcolsep}(}r@{)\hspace{2\tabcolsep}}
r@{.}l@{\hspace{0.5\tabcolsep}(}r@{)\hspace{2\tabcolsep}}
}
\multicolumn{22}{c}{$\beta = 2.4$, $L = 14$} \\
\hline \hline
\multicolumn{1}{c}{$s$} & 
\multicolumn{3}{l}{$a^2\sigma$} &	
\multicolumn{3}{l}{\% curr.} &
\multicolumn{3}{l}{$P$} &
\multicolumn{2}{l}{$l_{\max}$} &
\multicolumn{2}{l}{$l_{\sec}$} &
\multicolumn{3}{l}{$n_C$} &
\multicolumn{2}{l}{$l^\prime_{\sec}$}   &
\multicolumn{3}{l}{$n^\prime_C$} \\
\hline
0 & 	0&067$^{\dagger}$ & 2 & 	100&0 & 0 & 	56&8 &  3 &
	2441 & 141 &	121 & 48 &	249&3 & 5 &
			121 & 48 &	249&3 & 5 \\
1 & 	0&064 & 2 & 	 52&7 & 2 &	59&6 &  6 &
	1345 & 107 &	202 & 44 &	75&1 & 4 &
			73 & 22 &	61&8 & 4 \\
2 &	0&063 & 2 &  	 37&6 & 2 &	60&2 &  7 &
	965 & 90 &	216 & 41 &	34&2 & 2 &
			60 & 19 &	24&9 & 3 \\
3 &	0&062 & 2 &  	 30&3 & 2 &	61&0 &  8 &
	788 & 84 &	207 & 41 &	19&6 & 2 &
			51 & 15 &	12&1 & 2 \\
5 &	0&059 &	2 &	 22&9 & 1 &	61&8 & 10 &
	603 & 75 &	182 & 34 &	9&9 & 2 &
			39 & 12	&	5&4 & 1 \\
10 & 	0&055 & 2 &	 15&5 & 1 &	64&5 &  9 &
	423 & 50 &	136 & 24 &	4&9 & 1 &
			25 & 10 &	2&2 & 1 \\
\hline \hline
\end{tabular}
}
\end{center}
\caption{The evolution under $s$ monopole smoothing sweeps of; 
the string tension, 
the proportion of current remaining, the percolation parameter, the 
length of the two largest clusters and the number of clusters. Also 
given are the last two quantities when the largest cluster is excluded
from consideration. Results labelled $^{\dagger}$ are from $L=16$.}

\label{tab_huge_cool}
\end{table}

\begin{table}
\begin{center}
\begin{tabular}
{c*{2}{r@{.}l@{\hspace{0.5\tabcolsep}(}r@{)\hspace{2\tabcolsep}}}}
\hline \hline
&
\multicolumn{3}{c}{$s$} &
\multicolumn{3}{c}{$t$} \\
\hline
$(\beta = 2.3, L=12)$: &	-0&150	& 3 & 0&340	& 2 \\
$(\beta = 2.4, L=14)$: &	-0&175	& 3 & 0&350	& 1 \\
$(\beta = 2.5, L=16)$: & 	-0&191	& 2 & 0&355	& 1 \\
\hline \hline
\end{tabular}
\end{center}
\caption{Fitting $r_{\eff}(l) = s + t \surd l$ to the 
clusters.}
\label{tab_sqrt_fits}
\end{table}

\begin{table}
\begin{center}
\begin{tabular}
{l*{6}{r@{.}l@{\hspace{0.5\tabcolsep}(}r@{)\hspace{2\tabcolsep}}}}
\hline \hline
$L=16$: & 
\multicolumn{3}{c}{${\hbox{ln }} C$} &		
\multicolumn{3}{c}{$\gamma$} &		
\multicolumn{3}{c}{$c_p^1(\gamma)$} &
\multicolumn{3}{c}{$c_p^1(\gamma=3)$} &
\multicolumn{3}{c}{$c_p^2(\gamma)$} &
\multicolumn{3}{c}{$c_p^2(\gamma=3)$} \\
\hline
$\beta = 2.3$: & 10&38  & 21 & 
3&11 & 8 & 3&52 & 80 & 2&97 & 4 & 1&24 & 28 & 1&05 & 2 \\
$\beta = 2.4$: &  9&72 & 11 & 
2&90 & 4 & 4&45 &  60 & 5&20 & 25 & 1&14 & 15 & 1&34 & 4 \\
$\beta = 2.5$: &  9&30 & 12 & 
2&94 & 4 & 6&30 &  90 & 6&75 & 24 & 1&10 & 15 & 1&16 & 9 \\
\hline \hline
\end{tabular}
\end{center}
\caption{Power law fits and scaling behaviour of the smaller clusters, 
including the assumption that scaling is controlled by $\gamma=3$.}
\label{tab_pow_law}
\end{table}

\newpage

\end{document}